\documentclass[reprint,superscriptaddress,aps,prb]{revtex4-2}

\usepackage[utf8]{inputenc}

\usepackage{amssymb}
\usepackage{amsmath}
\usepackage{graphicx}
\usepackage{xcolor}
\usepackage[colorlinks=true, linkcolor=blue, citecolor=blue, urlcolor=blue]{hyperref}
\usepackage{bbold}
\usepackage{mathtools}
\usepackage{comment}
\usepackage[bbgreekl]{mathbbol}
\DeclareSymbolFontAlphabet{\amsmathbb}{AMSb}
\usepackage{dsfont}
\usepackage{cellspace}
\usepackage{scalerel}
\usepackage{bm}

\newcommand{\at}{\overset{\leftrightarrow}{\alpha}}
\newcommand{\Gt}{\overset{\leftrightarrow}{G}_0}
\newcommand{\nt}{\overset{\leftrightarrow}{n}}

\newcommand{\Bt}{\overset{\leftrightarrow}{B}}
\newcommand{\dt}{\overset{\leftrightarrow}{d}}

\newcommand{\vct}[1]{\mathbf{#1}}
\DeclareMathOperator{\Tr}{Tr}
\renewcommand\Re{\operatorname{Re}}
\renewcommand\Im{\operatorname{Im}}

\begin{document}

\title{Propulsion force and heat transfer for  nonreciprocal nanoparticles}

\author{Laila Henkes}
\affiliation{Institute for Theoretical Physics, Georg-August-Universit\"{a}t G\"{o}ttingen, 37077 G\"{o}ttingen, Germany}
\author{Kiryl Asheichyk}
\affiliation{Department of Theoretical Physics and Astrophysics, Belarusian State University, 5 Babruiskaya Street, 220006 Minsk, Belarus}
\author{Matthias Krüger}
\affiliation{Institute for Theoretical Physics, Georg-August-Universit\"{a}t G\"{o}ttingen, 37077 G\"{o}ttingen, Germany}

\begin{abstract}
We analyze heat transfer and Casimir forces involving a nonreciprocal nanoparticle. By dissecting the resulting expressions into reciprocal and nonreciprocal contributions, we find that the particle's self emission contains $++$ and $--$ terms, i.e., the particle's reciprocal ($+$) and  nonreciprocal ($-$) parts couple to the respective parts of its surrounding. In contrast, the heat transfer to the nanoparticle from the surrounding contains  $-+$ and $+-$  contributions, which we find to persist at equal temperatures. For two nanoparticles, such persistent transfer is found to require one particle to be nonreciprocal and the other to be anisotropic. The propulsion force for the nanoparticle, for which our results agree with previous work, is dominated by $\pm\mp$ terms, making it distinct from forces found for reciprocal particles. The amplitude of the propulsion force can be orders of magnitude larger than gravitational forces. Despite being distinct, we find the $\pm\mp$ terms to be bound by $\pm\pm$ terms, a consequence of passivity of the objects. For the force, this bound limits the efficiency in a heat engine setup, as observed for parallel plates before.
\end{abstract}

\maketitle

\section{Introduction}
\label{sec:Introduction}
The Casimir effect, an attractive force between two perfectly conducting plates at zero temperature, was first described in 1948~\cite{Casimir_original}. In 1956, this force was experimentally measured for the first time~\cite{Lif_measurement}, and the theory was expanded to dielectrics at nonzero temperature~\cite{Lif_temp}, where thermal fluctuations contribute. In thermal equilibrium, a free energy can be found, from which forces can be calculated~\cite{Emig_scattering, Rahi_expansionWaves_equilibrium, Eckhardt_correlator_eq_gf, Maia_CasimirEnergy, scatter_conv_eq}. Progress in experimental methods made it possible to measure these forces for different objects with high precision~\cite{exp1, exp2, exp3}.

With the development of fluctuational electrodynamics~\cite{Rytov_book,Rytov_paper}, calculation of the Casimir force and of radiative heat transfer out of thermal equilibrium became possible. This led to the scattering approach~\cite{Antezza_atom_plate, Messina2011, Kruger_correlator, TraceFormulas, Rodriguez_trace}. For systems with objects of simple geometries, like spheres, plates, or cylinders, analytical formulas for force and heat transfer were found~\cite{Polder_HT_plates, Narayanaswamy_HT_spheres, Antezza08, Antezza_atom_plate, Bimonte_plates, TraceFormulas, Golyk_cylinder, Kruger_spheres, Kruger_correlator}. Numerical methods have been developed, which allow for more general setups~\cite{Polimeridis_HeattransferNumerical, McCauley_numeric, Rodriguez_numericalArbitrary, Rodriguez_numerical, dipole_sphere1}. Several experiments on heat transfer have been performed~\cite{HT_exp1, HT_exp2, HT_exp3, HT_exp4}, nonequilibrium forces have been measured using an atomic cloud over a dielectric substrate~\cite{exp_neq}.

Allowing for optical nonreciprocity leads to interesting new phenomena~\cite{review_HT_1, review_HT_2, Biehs_review,Abdallah_Hall}, such as angular dependence of heat transfer \cite{Ekeroth_nanoparticles}. It has been shown that there is a nonzero ``persistent'' heat current for three nonreciprocal spheres at equal temperature~\cite{Zhu_epsilon_nonreci}, as well as propulsion forces, i.e., forces pointing in translationally invariant directions, of various types~\cite{Gelbwaser_nonreciprocalPlate, Khandekar_nonreciprocalPlate, gyrodparticle, Milton_propulsion_force}. While such forces seem to be more dominant in cases of nonreciprocity, they can also be found for reciprocal, anisotropic particles~\cite{Muller_ellipsoid}, and torques have been found as well~\cite{reid2017, strekha_eq, Milton_propulsion_force}. In Ref.~\cite{Gelbwaser_nonreciprocalPlate}, a general bound between propulsion force and heat transfer was derived, which is relevant for the efficiency of heat engine setups.

In this manuscript, we analyze heat transfer and propulsion forces specifically for nonreciprocal particles. We find that, for the self emission of the particle, the particle's nonreciprocity (described by the $ - $ term defined below) couples to the nonreciprocal part of the surrounding. In other words, in a reciprocal surrounding, the heat radiation of the particle is only sensitive to its reciprocal part (described by the $ + $ term). Furthermore, we find persistent heat current between two particles at equal temperature; this setup requires one particle to be nonreciprocal and the other one anisotropic. For the  propulsion force, the particle's nonreciprocity couples to the reciprocal part of the surrounding, but not to its nonreciprocal part. The magnitude of the propulsion force for a nonreciprocal particle near a planar surface can be large compared to its gravitational force. While force and heat transfer carry $\pm\mp$ terms that yield qualitatively different behavior due to nonreciprocity, these terms are bound in magnitude by $\pm\pm$ contributions. These bounds rely on the passivity of the media involved.
 
The paper is structured as follows: We give general formulas and discuss resulting properties for heat transfer in Sec.~\ref{sec:Heat_transfer}. In Sec.~\ref{sec:epsalpha}, we introduce the used dielectric properties of anisotropic or nonreciprocal nanoparticles. Resorting to the example of two nanoparticles, the self emission is analyzed in Subsec.~\ref{subsec:PHC_exampleSE}, and heat transfer in Subsec.~\ref{subsec:PHC_example}. In the latter, we show that there can be a nonzero heat transfer at equal temperature. The force acting on the nanoparticle is  introduced and analyzed in Sec.~\ref{sec:Force}. We  compute the lateral force explicitly for the example of a nonreciprocal particle in front of an isotropic plate (Sec.~\ref{sec:force_example}), finding that this force can be large compared to the gravitational force. Finally, we derive and discuss relations and bounds between heat transfer and lateral forces in Sec.~\ref{sec:Bound}. We conclude in Sec.~\ref{sec:concl}.

\section{General relations for heat transfer}
\label{sec:Heat_transfer}
\subsection{Recalling formulas for arbitrary objects}
\label{sec:RevisitingHTgeneral}
We consider two objects, described by permittivity and permeability tensors $ \bbespilon_i(\vct{r},\vct{r}';\omega) $ and $ \bbmu_i(\vct{r},\vct{r}';\omega) $ [$i=\{1,2\}$], which can depend on positions $ \vct{r} $, $ \vct{r}' $, and frequency $ \omega $. We introduce the corresponding potential \cite{Rahi_expansionWaves_equilibrium}, $ \amsmathbb{V}_i = \omega^2/c^2(\bbespilon_i-\amsmathbb{I}) + \nabla\!\times\! (\amsmathbb{I}-\bbmu_i^{-1})\nabla\times $, where $ \amsmathbb{I} = \mathcal{I}\delta^{(3)}(\vct{r}-\vct{r}') $ is the identity operator (with $ \mathcal{I} $ being the $ 3\times 3 $ identity matrix), and $ c $ is the speed of light in vacuum. Note that $ \amsmathbb{V}_i $ is only nonzero if both spatial arguments lie within object $ i $. For operator $ \amsmathbb{A}(\vct{r},\vct{r}';\omega) $ [below being potential, scattering operator, or Green's tensor], we define, 
\begin{equation}
\amsmathbb{A}_{\textrm{I}} = \frac{\amsmathbb{A}-\amsmathbb{A}^{\dagger}}{2i},
\label{eq:AI}
\end{equation}
where $\dagger$ implies the adjoint operator, i.e., transpose and complex conjugation. For $\amsmathbb{A}= A_{ij}(\vct{r},\vct{r}')$,  $ (A_{ij}(\vct{r},\vct{r}'))^T=A_{ji}(\vct{r}',\vct{r})$, i.e., the transpose involves matrix transpose  and permutation of spatial arguments. Note that care should be taken regarding the transpose in other representations, e.g., in plane wave basis \cite{Fan2020}. $\amsmathbb{A}_{\textrm{I}}$ is by construction Hermitian. We are in this manuscript interested in effects caused by nonreciprocity, recalling that for reciprocal systems, $ \amsmathbb{A}^{\dagger} = \amsmathbb{A}^* $ and $ \amsmathbb{A}_{\textrm{I}} = \textrm{Im}[\amsmathbb{A} ]$.

The radiative heat transfer, i.e., the heat emitted by object $ 1 $ at temperature $ T_1 $ and absorbed by object $ 2 $, is~\cite{soo, Kruger_conductivity}
\begin{equation}
H_1^{(2)} = \frac{2}{\pi}\int_0^{\infty}\!\! d\omega \Theta(\omega,T_1)\Tr\left\{\amsmathbb{V}_{1\textrm{I}}\amsmathbb{G}^{\dagger}\amsmathbb{V}_{2\textrm{I}}\amsmathbb{G}\right\},
\label{eq:HT12general}
\end{equation}
where $\Theta(\omega, T_1) = \hbar\omega\left[\exp\left(\frac{\hbar\omega}{k_{\textrm{B}}T_1}\right)-1\right]^{-1}$ is the mean energy of the emitted photons, with $ \hbar $ and $ k_{\textrm{B}} $ being Planck's and Boltzmann's constants, respectively. $\amsmathbb{G} $ is the Green's function (or tensor) of the two objects~\cite{TraceFormulas},
\begin{equation}
\amsmathbb{G}=(\amsmathbb{I}+\amsmathbb{G}_0\amsmathbb{T}_2)(\amsmathbb{I}-\amsmathbb{G}_0\amsmathbb{T}_1\amsmathbb{G}_0\amsmathbb{T}_2)^{-1}(\amsmathbb{I}+\amsmathbb{G}_0\amsmathbb{T}_1)\amsmathbb{G}_0,
\label{eq:GviaT}
\end{equation}
where $ \amsmathbb{G}_0 $ is the free Green's function. We introduced the scattering operators $ \amsmathbb{T}_i $, recalling~\cite{Rahi_expansionWaves_equilibrium, TraceFormulas},
\begin{equation}
\amsmathbb{T}_i = \amsmathbb{V}_i(\amsmathbb{I}-\amsmathbb{G}_0\amsmathbb{V}_i)^{-1}.
\label{eq:Tdef}
\end{equation}
In the above equations, operator notation is understood, i.e., operator multiplication includes integration over a joint coordinate and summation over a joint index~\cite{TraceFormulas}. When  spatial arguments are given explicitly, e.g., Eqs.~\eqref{eq:HT22PP_matrix} or~\eqref{eq:HTnet1PP2PP_antisymm} below, summation over indices is understood.

The heat radiation, or self emission, of object $ 2 $ at temperature $ T_2 $, i.e., the heat emitted by object $ 2 $ and (re)absorbed by itself in the presence of object $ 1 $, is given by~\footnote{We added a minus sign to $ H_2^{(2)}$ with respect to Ref.~\cite{TraceFormulas} for it to be nonnegative}~\cite{soo, Kruger_conductivity}
\begin{equation}
H_2^{(2)} = \frac{2}{\pi}\int_0^{\infty}\!\! d\omega \Theta(\omega,T_2)\Tr\left\{\amsmathbb{V}_{2\textrm{I}}\left(\amsmathbb{G}_{\textrm{I}}-\amsmathbb{G}^{\dagger}\amsmathbb{V}_{2\textrm{I}}\amsmathbb{G}\right)\right\}
\label{eq:HT22general}.
\end{equation}
$H_1^{(2)}\geq 0$ and $H_2^{(2)}\geq 0$ if $\amsmathbb{V}_{i\textrm{I}}\geq 0$, as assumed throughout in this manuscript, and expected for passive objects.

\subsection{Self emission of a small object: two distinct terms}
\label{subsec:HT2PP}
We aim to study self emission for the case where object $ 2 $ is small, and expand Eq.~\eqref{eq:HT22general} in linear order of the scattering operator of object $ 2 $, using Eqs.~\eqref{eq:Tdef} and~\eqref{eq:GviaT}.   
Keeping only terms linear in $ \amsmathbb{T}_2 $, we obtain
\begin{equation}
H_2^{(2)} = \frac{2}{\pi}\int_0^{\infty}\!\! d\omega \Theta(\omega,T_2)\Tr\left\{\amsmathbb{T}_{2\textrm{I}}\amsmathbb{G}_{1\textrm{I}}\right\},
\label{eq:HT22onescat}
\end{equation}
where $ \amsmathbb{G}_1= \amsmathbb{G}_0+\amsmathbb{G}_0\amsmathbb{T}_1\amsmathbb{G}_0$ is the Green's function with only object $ 1 $ present. For reciprocal objects, Eq.~\eqref{eq:HT22onescat} reproduces Eq.~(18) in Ref.~\cite{Asheichyk_heatradiation}.

Further insight is gained by decomposing operator $\amsmathbb{A}$ into  
symmetric and antisymmetric parts, indicated by the superscripts ``$+$'' and ``$-$'', respectively,
\begin{equation}
\amsmathbb{A}^{\pm} = \frac{1}{2}\left( \amsmathbb{A}\pm \amsmathbb{A}^T\right),
\label{eq:A_decompose}
\end{equation}
where $ T $ implies the matrix transpose and permutation of spatial arguments. For reciprocal objects, $\amsmathbb{T}$, $\amsmathbb{V}$ and $\amsmathbb{G}$ are symmetric, i.e., the antisymmetric parts vanish.

Using cyclic property of the trace, and $\Tr[\amsmathbb{A}]=\Tr[\amsmathbb{A}^T]$, it is straightforward to show that the trace of the product of a symmetric and an antisymmetric operator vanishes. We thus have
$\Tr\left\{\amsmathbb{T}^-_{2\textrm{I}}\amsmathbb{G}^+_{1\textrm{I}}\right\}=\Tr\left\{\amsmathbb{T}^+_{2\textrm{I}}\amsmathbb{G}^-_{1\textrm{I}}\right\}=0$. The self emission in Eq.~\eqref{eq:HT22onescat} is thus the sum of two terms,
\begin{equation}
H_2^{(2)} = \frac{2}{\pi}\int_0^{\infty}\!\! d\omega \Theta(\omega,T_2)\left[\Tr\left\{\amsmathbb{T}_{2\textrm{I}}^+\amsmathbb{G}^+_{1\textrm{I}}\right\}+\Tr\left\{\amsmathbb{T}_{2\textrm{I}}^-\amsmathbb{G}^-_{1\textrm{I}}\right\}\right].
\label{eq:HT22onescat_2}
\end{equation}
Equation~\eqref{eq:HT22onescat_2} states that, if the surrounding of particle~$ 2 $ is reciprocal, then only its reciprocal part contributes to the self emission. In other words, if the surrounding is reciprocal, the nonreciprocity of object $ 2 $ cannot be recognized in its self emission. The second term can only be finite if both the surrounding as well as object 2 are nonreciprocal. While the first term is nonnegative, i.e., 
\begin{align}
\Tr\left\{\amsmathbb{T}_{2\textrm{I}}^+\amsmathbb{G}^+_{1\textrm{I}}\right\} \geq 0,
\end{align}
the second one can have either sign, with the only apparent condition of
\begin{align}
\Tr\left\{\amsmathbb{T}_{2\textrm{I}}^-\amsmathbb{G}^-_{1\textrm{I}}\right\} \geq -\Tr\left\{\amsmathbb{T}_{2\textrm{I}}^+\amsmathbb{G}^+_{1\textrm{I}}\right\},
\end{align}
resulting from $\Tr\left\{\amsmathbb{T}_{2\textrm{I}}\amsmathbb{G}_{1\textrm{I}}\right\}\geq 0$. We will explicitly study the self emission for the configuration of two nonreciprocal particles in Subsec.~\ref{subsec:PHC_exampleSE} below.

\subsection{(Persistent) heat transfer for two small objects}
\label{subsec:PHC}
For heat transfer, it is insightful to expand  to linear order in both $\amsmathbb{T}_i$, yielding
\begin{equation}
H_1^{(2)} = \frac{2}{\pi}\int_0^{\infty}\!\! d\omega \Theta(\omega,T_1)\Tr\left\{\amsmathbb{T}_{1\textrm{I}}\amsmathbb{G}_0^{\dagger}\amsmathbb{T}_{2\textrm{I}}\amsmathbb{G}_0\right\}.
\label{eq:HT12onescat_1}
\end{equation}
The operator $\amsmathbb{G}_0^{\dagger}\amsmathbb{T}_{2\textrm{I}}\amsmathbb{G}_0$ has, in general, a symmetric and an antisymmetric part. Using that $\amsmathbb{T}_{2\textrm{I}}=\amsmathbb{T}_{2\textrm{I}}^\dagger$ and $\amsmathbb{G}_0^{\dagger}=\amsmathbb{G}_0^{*}$, we find
\begin{align}
\left(\amsmathbb{G}_0^{\dagger}\amsmathbb{T}_{2\textrm{I}}\amsmathbb{G}_0\right)^+&=\Re\left[\amsmathbb{G}_0^{\dagger}(\amsmathbb{T}_{2\textrm{I}}^++\amsmathbb{T}_{2\textrm{I}}^-)\amsmathbb{G}_0\right],\label{eq:+part}\\
\left(\amsmathbb{G}_0^{\dagger}\amsmathbb{T}_{2\textrm{I}}\amsmathbb{G}_0\right)^-&=i\Im\left[\amsmathbb{G}_0^{\dagger}(\amsmathbb{T}_{2\textrm{I}}^++\amsmathbb{T}_{2\textrm{I}}^-)\amsmathbb{G}_0\right],\label{eq:-part}
\end{align}
showing that heat transfer, in contrast to self emission, carries all combinations of $\pm$ contributions. It is furthermore interesting to note that
\begin{equation}
\Tr\left\{\amsmathbb{T}_{1\textrm{I}}^\pm\amsmathbb{G}_0^{\dagger}\amsmathbb{T}_{2\textrm{I}}^\pm\amsmathbb{G}_0\right\}=\Tr\left\{\amsmathbb{T}_{2\textrm{I}}^\pm\amsmathbb{G}_0^{\dagger}\amsmathbb{T}_{1\textrm{I}}^\pm\amsmathbb{G}_0\right\},
\label{eq:TG0TG0++--}
\end{equation}
showing that the $\pm\pm$ terms are symmetric (reciprocal)  in particle indices. Reciprocity in indices cannot, in general, be proven for heat transfer~\cite{Herz_2019}, and it is interesting to investigate the conditions for persistent heat transfer for two objects at equal temperature (recall that such current has been found for three objects~\cite{Zhu_epsilon_nonreci}), i.e., we consider the difference of $H_1^{(2)}(T)$  and $H_2^{(1)}(T)$,
\begin{align}
\notag & H^{1\to 2}(T) = H_1^{(2)}(T)-H_2^{(1)}(T) =\\ 
& \frac{2}{\pi}\int_0^{\infty}\!\! d\omega \Theta(\omega,T)\Tr\left\{\amsmathbb{T}_{1\textrm{I}}\amsmathbb{G}_0^{\dagger}\amsmathbb{T}_{2\textrm{I}}\amsmathbb{G}_0-\amsmathbb{T}_{2\textrm{I}}\amsmathbb{G}_0^{\dagger}\amsmathbb{T}_{1\textrm{I}}\amsmathbb{G}_0\right\}.
\label{eq:HTnet1PP2PP}
\end{align}
Decomposing $\amsmathbb{T}_{1}$ and $\amsmathbb{T}_{2}$ into symmetric and antiymmetric parts, according to Eq.~\eqref{eq:A_decompose}, yields
\begin{align}
\notag H^{1\to 2}(T) = & \ \frac{2}{\pi}\int_0^{\infty}\!\! d\omega \Theta(\omega,T)\\
& \times\Tr\left\{\amsmathbb{T}_{1\textrm{I}}^+\amsmathbb{G}_0^{\dagger}\amsmathbb{T}_{2\textrm{I}}^-\amsmathbb{G}_0-\amsmathbb{T}_{2\textrm{I}}^+\amsmathbb{G}_0^{\dagger}\amsmathbb{T}_{1\textrm{I}}^-\amsmathbb{G}_0\right\}.
\label{eq:HTnet1PP2PP_2}
\end{align}
This equation shows that, as indicated above, it is only the $\mp\pm$ terms that can contribute to $H^{1\to 2}(T)$. We will in Subsec.~\ref{subsec:PHC_example} provide more explicit formulas for persistent current to elucidate more requirements for it to occur.

\section{Electromagnetic response of nonreciprocal or anisotropic nanoparticles}
\label{sec:epsalpha}
To exemplify general quantities and properties discussed in the previous section, we consider nonmagnetic point particles for which the scattering operator $\amsmathbb{T}$ can be approximated in terms of the electric polarizability operator $\bbalpha$ (see Appendix~\ref{app:PP2})
\begin{align}
\notag \amsmathbb{T} &= 3\frac{\omega^2}{c^2}(\bbespilon-\amsmathbb{I})\left(\bbespilon+2\amsmathbb{I}\right)^{-1}= \frac{3}{R^3}\frac{\omega^2}{c^2}\bbalpha\\
&=\frac{3}{R^3}\frac{\omega^2}{c^2}\overset{\leftrightarrow}{\alpha}\delta^{(3)}(\vct{r}-\vct{r}'),
\label{eq:T2ss_alpha}
\end{align}
where $\overset{\leftrightarrow}{\alpha}$ and $R$ are the particle's polarizability matrix and radius, respectively. In the last step, we have assumed the local form for the permittivity tensor, $\bbespilon(\vct{r},\vct{r}';\omega) = \overset{\leftrightarrow}{\varepsilon}(\omega)\delta^{(3)}(\vct{r}-\vct{r}') $. We consider in this manuscript the matrix $\overset{\leftrightarrow}{\varepsilon}$ to be of the following form~\cite{Gelbwaser_nonreciprocalPlate, Gelbwaser_nonreciprocalEq, Zhu_epsilon_nonreci, Fan2020},
\begin{equation}
\overset{\leftrightarrow}{\varepsilon}=
\begin{pmatrix}
\varepsilon_{\textrm{p}} & 0 & 0\\
0 & \varepsilon_{\textrm{d}} & \varepsilon_{\textrm{s}} - i\varepsilon_{\textrm{f}}\\
0 & \varepsilon_{\textrm{s}} + i\varepsilon_{\textrm{f}} & \varepsilon_{\textrm{d}}
\end{pmatrix},
\label{eq:epsilon2}
\end{equation}
with all entries, in general, frequency dependent. For $\varepsilon_{\textrm{s}}=0$, Eq.~\eqref{eq:epsilon2} corresponds to a nonreciprocal medium, for example, a magneto-optical material with external magnetic field along the $x$ direction~\cite{Ishimaru} (see the sketch in Fig.~\ref{fig:setup_2spheres}, and Eq.~\eqref{eq:eps} for a specific example).

The case of $\varepsilon_{\textrm{f}}=0$ with $\varepsilon_{\textrm{s}}$ finite corresponds to a reciprocal, but anisotropic medium. Specifically, if $\varepsilon_{\textrm{s}}+\varepsilon_{\textrm{p}}=\varepsilon_{\textrm{d}}$, the given form can correspond to a uniaxial material with  optical axis  pointing along the bisecting line between $y$ and $z$ axes. 

The condition $\overset{\leftrightarrow}{\varepsilon}_\textrm{I}\geq 0$ yields simple conditions for its entries, which we will not spell out here.

For the given choice of $\overset{\leftrightarrow}{\varepsilon}$, the polarizability matrix $\overset{\leftrightarrow}{\alpha}$ takes the form
\begin{equation}
\overset{\leftrightarrow}{\alpha}=
\begin{pmatrix}
\alpha_{\textrm{p}} & 0 & 0\\
0 & \alpha_{\textrm{d}} & \alpha_{\textrm{s}} - i\alpha_{\textrm{f}}\\
0 & \alpha_{\textrm{s}} + i\alpha_{\textrm{f}} & \alpha_{\textrm{d}}
\end{pmatrix},
\label{eq:alpha2}
\end{equation}
with the entries [as functions of the entries of the permittivity matrix~\eqref{eq:epsilon2}] specified in Appendix~\ref{app:permittivityTensor}. They have the same physical interpretation as the entries of $ \overset{\leftrightarrow}{\varepsilon} $.

We note that the reciprocal version of $\overset{\leftrightarrow}{\alpha}$ with finite off-diagonal elements may also be found for a nonspherical object made of isotropic material~\cite{Absorption_and_scatteering, Muller_ellipsoid}.

\section{Examples: Heat transfer for two particles}
\label{sec:HTexamples}
\subsection{Self emission of a particle in the presence of another one}
\label{subsec:PHC_exampleSE}
\begin{figure}[!t]
\begin{center}
\includegraphics[width=0.45\textwidth]{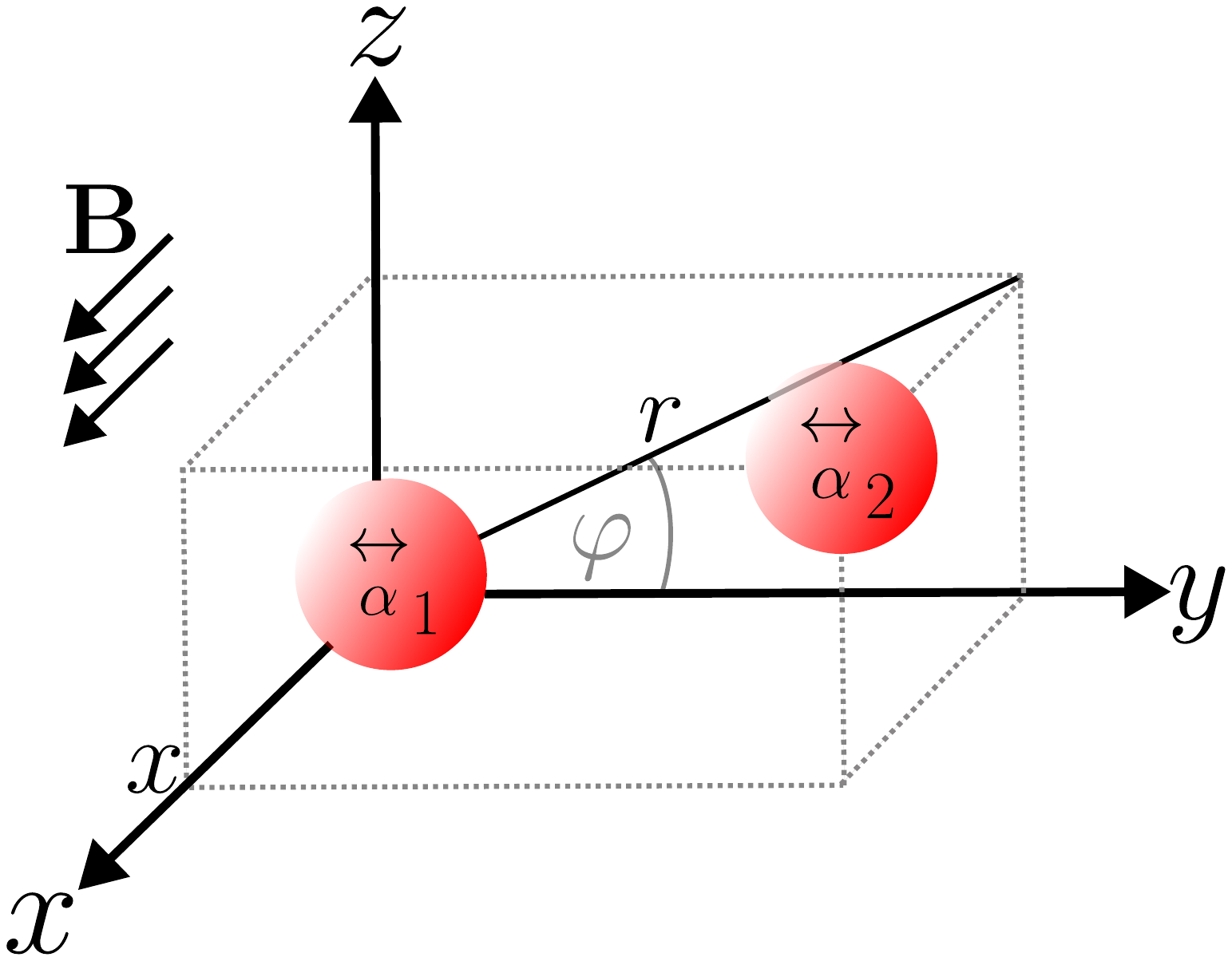}
\end{center}
\caption{\label{fig:setup_2spheres}The system of two spherical particles with polarizability matrices $\overset{\leftrightarrow}{\alpha}_{1} $ and $ \overset{\leftrightarrow}{\alpha}_{2}$, introducing the cylindrical coordinates ($r$, $\varphi$, $x$) to describe  the position of particle $ 2 $, with particle 1 at the origin. We denote ${\bf d}$ the vector connecting their centers, pointing at particle $ 2 $ (not shown for visibility), with $|{\bf d}| \equiv d$ being the interparticle distance. External magnetic field $ \vct{B} $ may give rise to antisymmetric parts of $ \overset{\leftrightarrow}{\alpha}_i$, making the particles optically nonreciprocal.} 
\end{figure}

Let us consider two particles placed at positions $ \vct{r}_1 = \vct{0} $ (i.e., at the origin) and $ \vct{r}_2 = (x,r\cos\varphi,r\sin\varphi)^T$, described by cylindrical coordinates $ (r,\varphi,x) $, with $ x $ being the reference axis, see Fig.~\ref{fig:setup_2spheres}. We denote the interparticle vector $ \vct{d} \equiv \vct{r}_2 $, i.e., their distance is $d=|\vct{r}_2|=\sqrt{r^2+x^2}$. In this setup, using operator~\eqref{eq:T2ss_alpha} in Eq.~\eqref{eq:HT22onescat} and assuming that the Green's function hardly varies within the volumes of the particles, we get for the self emission of particle $ 2 $
\begin{equation}
H_2^{(2)} = \frac{8}{c^2}\int_0^{\infty}\!\! d\omega \Theta(\omega,T_2)\omega^2\Tr\left\{\overset{\leftrightarrow}{\alpha}_{2\textrm{I}}\amsmathbb{G}_{1\textrm{I}}(\vct{r}_2,\vct{r}_2)\right\}.
\label{eq:HT22PP_matrix}
\end{equation}
As discussed in Sec.~\ref{subsec:HT2PP}, there are two distinct contributions, coupling symmetric  and antisymmetric parts of the particles' polarizabilities, respectively, such that Eq.~\eqref{eq:HT22PP_matrix} can be written as
\begin{align}
\notag H_2^{(2)} = & \ \frac{8}{c^2}\int_0^{\infty}\!\! d\omega \Theta(\omega,T_2)\omega^2\\
& \times \left[\Tr\left\{\overset{\leftrightarrow}{\alpha}_{2\textrm{I}}^+\amsmathbb{G}_{1\textrm{I}}^+(\vct{r}_2,\vct{r}_2)\right\} + \Tr\left\{\overset{\leftrightarrow}{\alpha}_{2\textrm{I}}^-\amsmathbb{G}_{1\textrm{I}}^-(\vct{r}_2,\vct{r}_2)\right\}\right].
\label{eq:HT22PP_matrix_decompose}
\end{align}
Equations~\eqref{eq:HT22PP_matrix} and~\eqref{eq:HT22PP_matrix_decompose} are so far valid for an arbitrary object $ 1 $, and we now turn to a point particle $ 1 $ with the Green's tensor \cite{Asheichyk_heatradiation}
\begin{equation}
\amsmathbb{G}_1(\vct{r},\vct{r}') = \amsmathbb{G}_0(\vct{r},\vct{r}') + 4\pi\frac{\omega^2}{c^2}\amsmathbb{G}_0(\vct{r},\vct{r}_1)\overset{\leftrightarrow}{\alpha}_1\amsmathbb{G}_0(\vct{r}_1,\vct{r}'),
\label{eq:G1PP}
\end{equation}
where the traces are evaluated in Appendix~\ref{app:traces_self_emission}.

Note that the vacuum part [stemming form $ \amsmathbb{G}_0 $ in Eq.~\eqref{eq:G1PP}] of the self emission in Eqs.~\eqref{eq:HT22PP_matrix} or~\eqref{eq:HT22PP_matrix_decompose} is proportional to the volume of particle $ 2 $, whereas the interaction part [stemming form the second term in Eq.~\eqref{eq:G1PP}] is proportional to the product of particle volumes. 

In this subsection (i.e., for self emission), we consider the particles to be isotropic ($ \alpha_{i\textrm{s}} = 0 $); the generalization to anisotropic particles is given in Appendix~\ref{app:traces_self_emission}. We start by reminding the reader of the self emission for the case of reciprocal media.

\subsubsection{Reciprocal  particles}
\label{par:SE_rec_iso}
For reciprocal particles ($ \alpha_{i\textrm{f}} = 0 $, $ \alpha_{i\textrm{d}} = \alpha_{i\textrm{p}} \equiv \alpha_i$), only the $++$ term in Eq.~\eqref{eq:HT22PP_matrix_decompose} is finite.  
The self emission becomes
\begin{align}
\notag & \underset{\textrm{rec}}{H_2^{(2)}} = \underset{\textrm{rec}}{H_{2,\textrm{vac}}^{(2)}} + \frac{4}{\pi d^6}\int_0^{\infty}\!\! d\omega \Theta(\omega,T_2)\Im[\alpha_{2}]\\
& \times \Im\left[\alpha_{1}e^{2i\frac{\omega}{c}d}\left(3-6i\frac{\omega}{c}d-5\frac{\omega^2}{c^2}d^2+2i\frac{\omega^3}{c^3}d^3+\frac{\omega^4}{c^4}d^4\right)\right],
\label{eq:HT2PP2PP_recipiso}
\end{align}
where the vacuum part (i.e., without particle $ 1 $ present) is~\cite{Asheichyk_heatradiation}
\begin{equation}
\underset{\textrm{rec}}{H_{2,\textrm{vac}}^{(2)}} = \frac{4}{\pi c^3}\int_0^{\infty}\!\! d\omega \Theta(\omega,T_2)\omega^3\Im[\alpha_{2}].
\label{eq:HT2PP2PP_recipiso_vac}
\end{equation}
The effect of particle $ 1 $ in Eq.~\eqref{eq:HT2PP2PP_recipiso} is insignificant for the far field, i.e., large $ d $, but can be significant in the near field, where $ d \ll \lambda_{T_2} $ [with $ \lambda_{T_2} = \hbar c/(k_{\textrm{B}}T_2) $ being the thermal wavelength], see Fig.~\ref{fig:SEddep}. In this limit,
\begin{align}
\notag \lim_{d\ll\lambda_{T_2}}\underset{\textrm{rec}}{H_2^{(2)}} = & \ \underset{\textrm{rec}}{H_{2,\textrm{vac}}^{(2)}} + \frac{4}{\pi}\int_0^{\infty}\!\! d\omega \Theta(\omega,T_2)\Im[\alpha_{2}]\\
& \times \left\{\Im[\alpha_{1}]\frac{3}{d^6}+\Re[\alpha_{1}]\frac{22}{15d}\frac{\omega^5}{c^5}\right\}.
\label{eq:HT2PP2PP_recipiso_NF}
\end{align}
The first term in brackets in Eq.~\eqref{eq:HT2PP2PP_recipiso_NF}, scaling as $ d^{-6} $ and being positive, corresponds to the heat transfer from particle $ 2 $ to particle $ 1 $ (compare to, e.g., Eq.~(23) in Ref.~\cite{Asheichyk_heatradiation}). The second term, which scales as $ d^{-1} $, and which is the leading term for small $d$ for a nonabsorbing particle $ 1 $, can be of either sign, and it describes the effect of particle $ 1 $ on the heat transfer from particle $ 2 $ to the environment.

\subsubsection{Nonreciprocal particles: $ \vct{B} $ and $ \vct{d} $ parallel}
\label{par:SE_iso_par}
We assume nonreciprocity to occur because of an external magnetic field $ \vct{B} $ pointing along the $x$ axis (see Fig.~\ref{fig:setup_2spheres}), which gives rise to the polarizability tensor in Eq.~\eqref{eq:alpha2}, with $\alpha_\textrm{f}$ finite and $\alpha_\textrm{s}=0$ for both particles. In this case, the self emission depends on the relative angle between $ \vct{B} $ and $ \vct{d} $, as was also observed for the heat transfer in Ref.~\cite{Ekeroth_nanoparticles}. For small $B$, as shown in Appendix~\ref{app:tensorial}, it carries an isotropic term $\sim B^2$ and a term $\sim ({\bf d}\times {\bf B} )\cdot ({\bf d}\times {\bf B})$, reminiscent of the dependence found for the heat transfer~\cite{Ekeroth_nanoparticles}. For symmetry reasons, the self emission does not depend on the azimuthal angle $ \varphi $ (see Fig.~\ref{fig:setup_2spheres}). 

We start with the case where particle 2 is placed on the $x$ axis, i.e., $ \vct{d} \ || \ \vct{B} $ or ${\bf d}\times {\bf B}={\bf 0}$, probing the isotropic term. Using Eqs.~\eqref{eq:HT22PP_matrix_decompose} and~\eqref{eq:TrSE}, we find

\begin{align}
\notag \underset{\parallel B}{H_2^{(2)}} = H_{2,\textrm{vac}}^{(2)} + \ & \frac{2}{\pi d^6}\int_0^{\infty}\!\! d\omega \Theta(\omega,T_2)\\
& \times\!\!\!\! \sum_{m=\{\textrm{p},\textrm{d},\textrm{f}\}}\!\!\!\!\Im[\alpha_{2m}]\Im\left[\alpha_{1m}g_m^{\parallel}\left(\frac{\omega}{c}d\right)\right],
\label{eq:HT2PP2PP_iso_parB}
\end{align}
where the functions $g_m^{\parallel}$ are given by
\begin{subequations}
\begin{align}
g_{\textrm{p}}^{\parallel}(x)&= 4e^{2ix}\left(1-2ix-x^2\right),\label{eq:g_par_p}\\
g_{\textrm{d}}^{\parallel}(x)&= g_{\textrm{f}}^\parallel(x)= 2e^{2ix}\left(1-2ix-3x^2+2ix^3+x^4\right).\label{eq:g_par_df}
\end{align}
\end{subequations}
The self emission of particle $ 2 $ in vacuum reads, 
\begin{align}
H_{2,\textrm{vac}}^{(2)}= \frac{4}{3\pi c^3}\int_0^\infty d\omega \Theta(\omega,T_2)\omega^3\Im[\alpha_{2\textrm{p}}+2\alpha_{2\textrm{d}}],
\label{eq:HT22vac}
\end{align}
which, in the absence of the magnetic field, reduces to Eq.~\eqref{eq:HT2PP2PP_recipiso_vac}.
For $d\ll\lambda_{T_2}$, Eq.~\eqref{eq:HT2PP2PP_iso_parB} reduces to
\begin{align}
\notag & \lim_{d\ll\lambda_{T_2}}\underset{\parallel B}{H_2^{(2)}} =H_{2,\textrm{vac}}^{(2)} +  \frac{4}{\pi}\int_0^{\infty}\!\! d\omega \Theta(\omega,T_2)\bigg\{\frac{1}{d^6}\big(2\Im[\alpha_{2\textrm{p}}]\\
\notag & \times \Im[\alpha_{1\textrm{p}}]+\Im[\alpha_{2\textrm{d}}]\Im[\alpha_{1\textrm{d}}]+\Im[\alpha_{2\textrm{f}}]\Im[\alpha_{1\textrm{f}}]\big)+\frac{4}{3d^3}\frac{\omega^3}{c^3}\\
& \times \!\big(\!\Im[\alpha_{2\textrm{p}}]\Re[\alpha_{1\textrm{p}}]-\Im[\alpha_{2\textrm{d}}]\Re[\alpha_{1\textrm{d}}]-\Im[\alpha_{2\textrm{f}}]\Re[\alpha_{1\textrm{f}}]\big)\!\bigg\}.
\label{eq:HT2PP2PP_iso_parB_NF}
\end{align}
In contrast to the self emission for reciprocal particles in Eq.~\eqref{eq:HT2PP2PP_recipiso_NF}, an additional power law $ d^{-3} $ appears in Eq.~\eqref{eq:HT2PP2PP_iso_parB_NF}. As above, the term $ d^{-6} $ corresponds to the heat transfer from particle $ 2 $ to particle $ 1 $, while the term $ d^{-3} $ describes heat transfer to the environment, stimulated by particle $ 2 $ without absorbing itself. Due to the different power law, this term, at small $d$, is larger  compared to Eq.~\eqref{eq:HT2PP2PP_recipiso_NF}. Interestingly, it appears due to the diagonal entries of the polarizabilities not being equal, and it also obtains a contribution from nonreciprocal parts.

The diagonal entries of the polarizabilities in Eq.~\eqref{eq:HT2PP2PP_iso_parB}, i.e., $m=\{\textrm{p},\textrm{d}\}$ correspond to the $++$ term in Eq.~\eqref{eq:HT22PP_matrix_decompose}, while  $m={\textrm{f}}$ corresponds to the $--$ term. We remind that, if $\at_{i\textrm{I}}\geq 0$, the self emission is positive, the term $++$ is positive, while the term $--$ can have either sign. Notably, $\at_{i\textrm{I}}\geq 0$ requires $\left|\Im[\alpha_{i\textrm{f}}]\right|\leq \Im[\alpha_{i\textrm{d}}]$, showing that $\Im[\alpha_{i\textrm{f}}]$ and the $--$ term can have either sign.

\subsubsection{Nonreciprocal particles: $ \vct{B} $ and $ \vct{d} $ perpendicular}
\label{par:SE_iso_perp}
For the perpendicular alignment, the terms $\sim B^2$ and $\sim ({\bf d}\times {\bf B} )\cdot ({\bf d}\times {\bf B})$ contribute. We find in this case the same form as Eq.~\eqref{eq:HT2PP2PP_iso_parB}, with the functions $g_m^{\parallel}$ replaced by
\begin{subequations}
\begin{align}
g_{\textrm{p}}^{\perp}(x)&=e^{2ix}\left(1-2ix-3x^2+2ix^3+x^4\right),\label{eq:g_perp_p}\\
g_{\textrm{d}}^\perp(x)&= e^{2ix}\left(5-10ix-7x^2+2ix^3+x^4\right),\label{eq:g_perp_d}\\
g_{\textrm{f}}^\perp(x)&=4e^{2ix}\left(-1+2ix+2x^2-ix^3\right).\label{eq:g_perp_f}
\end{align}
\end{subequations}
The near-field limit for the perpendicular case reads as
\begin{align}
\notag & \lim_{d\ll\lambda_{T_2}}\underset{\perp B}{H_2^{(2)}} =H_{2,\textrm{vac}}^{(2)} +  \frac{4}{\pi}\int_0^{\infty}\!\! d\omega \Theta(\omega,T_2)\bigg\{\frac{1}{2d^6}\big(\Im[\alpha_{2\textrm{p}}]\\
\notag & \times \Im[\alpha_{1\textrm{p}}]+5\Im[\alpha_{2\textrm{d}}]\Im[\alpha_{1\textrm{d}}]-4\Im[\alpha_{2\textrm{f}}]\Im[\alpha_{1\textrm{f}}]\big)\\
\notag & \ \ \ -\frac{2}{3d^3}\frac{\omega^3}{c^3}\big(\!\Im[\alpha_{2\textrm{p}}]\Re[\alpha_{1\textrm{p}}]-\Im[\alpha_{2\textrm{d}}]\Re[\alpha_{1\textrm{d}}]\\
& \ \ \ \ \ \ \ \ \ \ \ \ \ \ \ \ \ \ \ \ \ \ \ \ \ \ \ \ \ \ \ \ \ \ \ \ \ -\Im[\alpha_{2\textrm{f}}]\Re[\alpha_{1\textrm{f}}]\big)\!\bigg\}.
\label{eq:HT2PP2PP_iso_perpB_NF}
\end{align}
As for the parallel case, the $++$ and $--$ contributions are separated, as dictated by Eq.~\eqref{eq:HT22PP_matrix_decompose}. Here, the term $--$ carries an overall minus sign, i.e., the leading term for small $d$ is negative for particles made of same material,  see Eq.~\eqref{eq:HT2PP2PP_iso_perpB_NF}. 

Notably, comparing the cases of parallel and perpendicular, we conclude that, for the special configuration with angle such that  $({\bf d}\times {\bf B} )\cdot ({\bf d}\times {\bf B})=\frac{B^2 d^2}{3}$, the leading term for small $d$ of the $--$ term vanishes.

\subsubsection{Numerical example}
\label{par:SE_iso_num}
\begin{figure*}[!t]
\begin{center}
\includegraphics[width=0.8\linewidth]{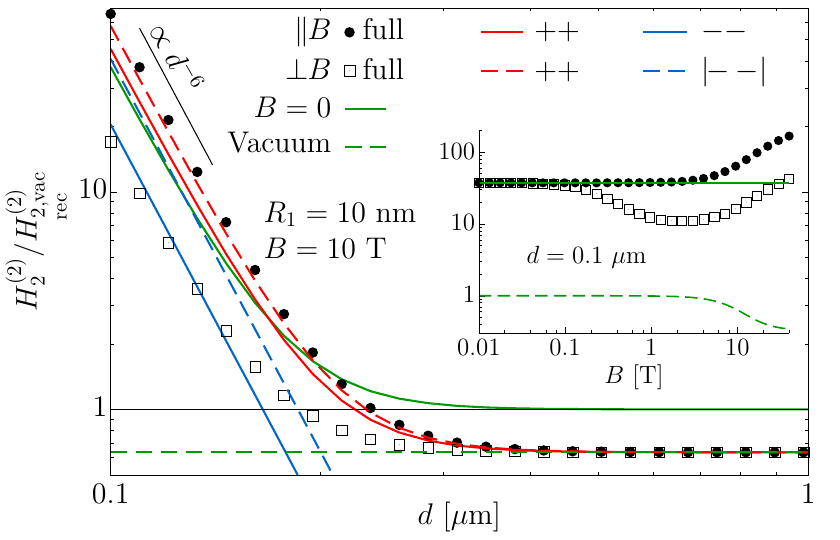}
\end{center}
\caption{\label{fig:SEddep}
Two nonreciprocal particles, modelled by $n$-doped InSb in an external magnetic field $ B = 10 \ \textrm{T} $,  with interparticle distance $ d $. Shown is the heat radiation of particle 2  at temperature $ T_2 = 300 \ \textrm{K} $, normalized by the result in the absence of particle $ 1 $  and $B=0$. $ R_1 = 10 \ \textrm{nm} $. Shown are the cases of $ {\bf B} $ parallel ($ \parallel \!\! B $) and perpendicular ($ \perp \!\! B $) to the interparticle vector ${\bf d}$, as well as the cases of zero field  ($ B = 0 $) and the heat radiation without particle $ 1 $ (``Vacuum''). Also shown are the individual ($ ++ $) and  ($ -- $) parts, as labeled.  Solid line at a value of $ 1 $ is included as a guide to the eye. Inset: Dependence on the magnetic field for $ d = 0.1 \ \mu\textrm{m} $, where the solid and dashed lines correspond to no field and no particle $ 1 $, respectively.}
\end{figure*}

Figure~\ref{fig:SEddep} shows the self emission for parallel and perpendicular cases using Eqs.~\eqref{eq:eps} to model the dielectric tensor, corresponding to $n$-doped InSb particles. As the reference situation for normalization, we choose particle~$ 2 $ in isolation and with magnetic field strength $ B=0 $, i.e., Eq.~\eqref{eq:HT2PP2PP_recipiso_vac}. Normalized in this way, the self emission does not depend on the size of particle $ 2 $, but it depends on the size of particle $ 1 $ (if present), which we fix to  $ R_1 = 10 \ \textrm{nm} $.

The main plot of Fig.~\ref{fig:SEddep} shows the self emission as a function of $ d $ for fixed $ B = 10 \ \textrm{T} $, alongside with the result for $ B = 0 $ (no magnetic field). The self emission shows the expected increase when $d$ is sub-micron, ultimately scaling as $d^{-6}$. In this limit, the finite magnetic field $ B $ either enhances or reduces the particle emission: It is enhanced for the parallel case and reduced for the perpendicular. This is because, in this limit, both $++$ and $--$ terms are positive for the parallel case, while the $--$ term is negative for the perpendicular case [compare Eqs.~\eqref{eq:HT2PP2PP_iso_parB_NF} and~\eqref{eq:HT2PP2PP_iso_perpB_NF}]. For large $d$, the self emission, as expected, converges to the result in isolation with or without (the reference case) $ B $ present.

In the inset of Fig.~\ref{fig:SEddep}, we show the dependence on $ B $. If particle $ 1 $ is absent, the  self emission is suppressed by the field $B$,  changing  noticeable around $ B \approx 2 \ \textrm{T} $, and being  roughly a factor of $ 1.5 $ smaller when $ B = 10 \ \textrm{T} $. In agreement with the main plot, the $ B $-dependence in the presence of particle $ 1 $ is different for the parallel and perpendicular alignment. For the former case, the heat radiation increases as $ B $ increases. For the latter case, the radiation is suppressed, and the dependence on $ B $ is nonmonotonic, featuring a minimum at $ B \approx 2.3 \ \textrm{T} $. 

The self emission is an even function of $B$, and, for small $B$, all curves are of order $B^2$, compare Eqs.~\eqref{eq:BtensorialSE}. Notably, terms linear in $B$ are absent due to the absence of  $\pm\mp$ terms in  Eq.~\eqref{eq:HT22PP_matrix_decompose}.

\subsection{Persistent heat transfer for two particles in thermal equilibrium}
\label{subsec:PHC_example}
As mentioned in Subsec.~\ref{subsec:HT2PP}, Eq.~\eqref{eq:HTnet1PP2PP_2} allows, in principle, for a finite persistent current for two particles. We aim to study it more explicitly for the case of two point particles. In this limit, Eq.~\eqref{eq:HTnet1PP2PP_2} reads 
\begin{align}
\notag H^{1\to 2}(T) & = \frac{64\pi}{c^4}\int_0^{\infty}\!\! d\omega \Theta(\omega,T)\omega^4\\
\notag & \times \Bigg[\Tr\left\{\overset{\leftrightarrow}{\alpha}_{1\textrm{I}}^+\amsmathbb{G}_0^*(\vct{r}_1,\vct{r}_2)\overset{\leftrightarrow}{\alpha}_{2\textrm{I}}^-\amsmathbb{G}_0(\vct{r}_1,\vct{r}_2)\right\}\\
& \ \ - \Tr\left\{\overset{\leftrightarrow}{\alpha}_{2\textrm{I}}^+\amsmathbb{G}_0^*(\vct{r}_1,\vct{r}_2)\overset{\leftrightarrow}{\alpha}_{1\textrm{I}}^-\amsmathbb{G}_0(\vct{r}_1,\vct{r}_2)\right\}\Bigg],
\label{eq:HTnet1PP2PP_antisymm}
\end{align}
which is proportional to the volumes of the two particles. 
As noted in Eq.~\eqref{eq:HTnet1PP2PP_2}, only the $\pm\mp$ terms contribute. Notably, for $\at_1=\at_2$, $H^{1\to 2}$ vanishes.

While it is expected that one of the particles must be nonreciprocal, it is surprising that the other particle must be anisotropic. This can be seen explicitly by simplifying Eq.~\eqref{eq:HTnet1PP2PP_antisymm} further for the special choice of Eq.~\eqref{eq:alpha2}, where we find
\begin{align}
\notag H^{1\to 2}(T) = & \ \frac{16 r^2\cos(2\varphi)}{\pi c^3d^5}\int_0^{\infty}\!\! d\omega \Theta(\omega,T)\omega^3\\
& \times \left(\Im[\alpha_{1\textrm{s}}]\Im[\alpha_{2\textrm{f}}] - \Im[\alpha_{2\textrm{s}}]\Im[\alpha_{1\textrm{f}}]\right).
\label{eq:HTnet1PP2PP_explicit}
\end{align}
Recall, from Fig.~\ref{fig:setup_2spheres}, $ r $ and $ \varphi $ are the radial and azimuthal coordinates of particle $ 2 $.

We note that $ H^{1\to 2}(T) $ in Eq.~\eqref{eq:HTnet1PP2PP_explicit} is proportional to $ \cos(2\varphi) $, its amplitude and sign changes with $\varphi$, i.e., when precessing $ \vct{d} $ around $ \vct{B} $. $ |H^{1\to 2}(T)| $ is maximal for $\varphi=\{0,\pi/2,\pi,3\pi/2\}$) and it vanishes for $\varphi=\{\pi/4,3\pi/4,5\pi/4,7\pi/4\}$.

Due to the factor $r^2$, the current also vanishes when $ \vct{d} \ || \ \vct{B} $, i.e., the particles are aligned with the magnetic field. 

$ H^{1\to 2}(T) $, resulting from $\pm\mp$ contributions, is linear in the magnetic field for small $ B $, and it thus changes sign with changing the direction of $\vct{B}$. These observations agree with a general structure of the transfer in Eq.~\eqref{eq:ndBI}, $~({\bf n}\cdot {\bf d}) (\vct{n}\times{\bf B})\cdot \vct{d} $, with $ \vct{n} $  the optical axis of the anisotropic medium.

While it is well understood that there can be a net heat flow in a system involving at least three nonreciprocal bodies at the same temperatures~\cite{Zhu_epsilon_nonreci}, Eq.~\eqref{eq:HTnet1PP2PP_explicit} implies such a flow between $ 2 $ bodies, which to our knowledge was not observed previously.

With all temperatures equal, net radiation from any particle must vanish~\cite{Gelbwaser_nonreciprocalPlate}. For the observed persistent transfer for two particles, the particles must thus exchange energy with the environment. If $H^{1\to 2}$ is positive, there must be positive energy transfer from particle~$ 2 $ to the environment, and there must be be positive energy transfer from the environment to particle~$ 1 $. Although not computed explicitly here, they must equal in magnitude the result in Eq.~\eqref{eq:HTnet1PP2PP_explicit}.

Eq.~\eqref{eq:HTnet1PP2PP_explicit} carries the temperature dependent distribution $\Theta(\omega,T)$ as a pre-factor. One may however also attribute such persistent transfer for zero point fluctuations. This has to be investigated in future work.

\section{Force on a nonreciprocal  particle: General formulas and properties}
\label{sec:Force}
\begin{figure}[!t]
\begin{center}
\includegraphics[width=0.45\textwidth]{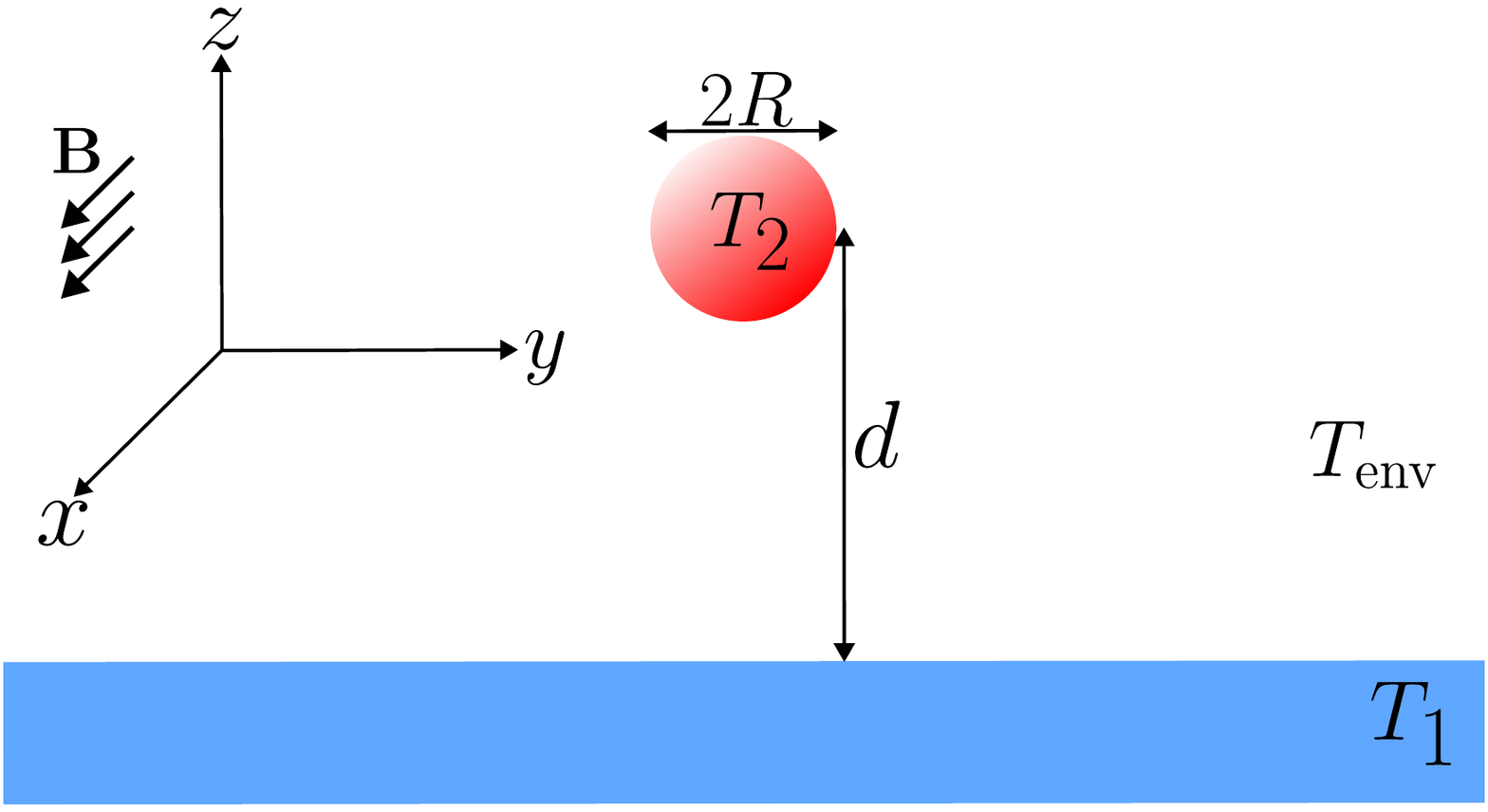}
\end{center}
\caption{\label{fig:setup}A reciprocal isotropic semi-infinite (occupying the half-space $ z \leq 0 $) plate at temperature $T_1$ and a small sphere made of magneto-optical material at temperature $T_2$ with radius $R$ at distance $d$ to the plate. The environment is at temperature $T_\textrm{env}$. Due to the external magnetic field $\vct{B}$ in the $x$ direction, the sphere is optically nonreciprocal.}
\end{figure}

We consider object $ 2 $ opposite object $ 1 $ which is translationally invariant in direction $y$, see the sketch in Fig.~\ref{fig:setup} for the case of a sphere opposite a plate. We are interested in the force acting on object $ 2 $ pointing along the $y$ axis, i.e., a lateral, or propulsion, force~\cite{Muller_ellipsoid, Gelbwaser_nonreciprocalPlate, Khandekar_nonreciprocalPlate, Milton_lateralforce_eq, Milton_propulsion_force}. In this setup, this force vanishes at thermal equilibrium~\cite{Gelbwaser_nonreciprocalEq}. It also vanishes in setups with spatial symmetry, thus relying on anisotropy or nonreciprocity. For different temperatures of the objects ($ T_1 $ and $ T_2 $) and the environment ($ T_\textrm{env} $), it can be written as~\cite{TraceFormulas}
\begin{align}
\notag F_{y}^{(2)}(T_1,T_2,T_{\textrm{env}})= & \ F^{(2)}_{1,y}(T_1)+F^{(2)}_{2,y}(T_2)\\
&-F^{(2)}_{1,y}(T_\textrm{env})-F^{(2)}_{2,y}(T_\textrm{env}),
\label{eq:total_force}
\end{align}
where $F_{1,y}^{(2)}$ is the interaction force, arising from fluctuations inside object $1$, and $F_{2,y}^{(2)}$ is the self force due to fluctuations inside object $2$~\cite{TraceFormulas}. The last two terms of Eq.~\eqref{eq:total_force} allow for a finite environment temperature, and Eq.~\eqref{eq:total_force} uses that the force vanises when all temperatures are equal~\cite{TraceFormulas}.

We adapt the methods of Refs.~\cite{TraceFormulas, Gelbwaser_nonreciprocalPlate} to find the forces in linear order in $\amsmathbb{T}_2$. The self force is
\begin{align}
\notag & F_{2,y}^{(2)}(T_2)=\frac{2}{\pi }\int_0^\infty\! \frac{d\omega}{\omega} \Theta(\omega,T_2)\textrm{Tr}\left\{i\partial_{y}\amsmathbb{T}_{2\textrm{I}}\amsmathbb{G}_{1\textrm{I}}\right\},\\
&=\frac{2}{\pi }\int_0^\infty\! \frac{d\omega}{\omega} \Theta(\omega,T_2)\textrm{Tr}\left\{i\partial_{y}\amsmathbb{T}_{2\textrm{I}}^+\amsmathbb{G}_{1\textrm{I}}^-+i\partial_{y}\amsmathbb{T}_{2\textrm{I}}^-\amsmathbb{G}_{1\textrm{I}}^+\right\}.
\label{eq:F_ss_trace_symmery}
\end{align}
$ \partial_y $ is the derivative with respect to $y$. In the second line, we used that $\partial_y$ is antisymmetric and that it commutes with $\amsmathbb{G}_{1}$ due to translational invariance. The second line shows that only the $\pm\mp$ terms contribute to the self force, similar to the persistent heat current in Eq.~\eqref{eq:HTnet1PP2PP_antisymm} and in contrast to the self emission in Eq.~\eqref{eq:HT22PP_matrix_decompose}. This implies that the self force, in the first order of $ \amsmathbb{T}_2 $, requires nonreciprocity, in agreement with the previous findings that a propulsion force for reciprocal anisotropic objects appears only in the second order~\cite{Muller_ellipsoid}. Equation~\eqref{eq:F_ss_trace_symmery} also states that there are two possible scenarios: A nonreciprocal particle with a reciprocal surrounding, as studied in Ref.~\cite{Milton_propulsion_force} and as considered below, and a reciprocal particle in a nonreciprocal surrrounding, as studied in Ref.~\cite{Khandekar_nonreciprocalPlate}. There is no additional term when both of them are nonreciprocal. 

The interaction force reads
\begin{align}
\notag & F_{1,y}^{(2)}(T_1)=\frac{2}{\pi }\int_0^\infty\! \frac{d\omega}{\omega} \Theta(\omega,T_1)   \textrm{Tr}\left\{i\partial_y\amsmathbb{T}_{2\textrm{I}}\amsmathbb{G}_1\amsmathbb{V}_{1\textrm{I}}\amsmathbb{G}_1^\dagger\right\}\\
\notag & =\frac{2}{\pi }\int_0^\infty\! \frac{d\omega}{\omega} \Theta(\omega,T_1)\bigg[\textrm{Tr}\left\{i\partial_y\amsmathbb{T}_{2\textrm{I}}^{-}\left(\amsmathbb{G}_1\amsmathbb{V}_{1\textrm{I}}\amsmathbb{G}_1^\dagger\right)^+\right\}\\
& \ \ \ \ +\textrm{Tr}\left\{i\partial_y\amsmathbb{T}_{2\textrm{I}}^{+}\left(\amsmathbb{G}_1\amsmathbb{V}_{1\textrm{I}}\amsmathbb{G}_1^\dagger\right)^-
\right\}\bigg].
\label{eq:Fps_sym}
\end{align}
In the last equality, we have split the force into products of symmetric and antisymmetric operators. This shows that also the interaction force is a sum of two distinct terms. However, the antisymmetric part of  $\amsmathbb{G}_1\amsmathbb{V}_{1\textrm{I}}\amsmathbb{G}_1^\dagger$ does not require nonreciprocity of object $ 1 $, so that this force can also exist for two reciprocal objects, as seen in the second term of Eq.~\eqref{eq:F_y_firstorder} below. Compare also Eqs.~\eqref{eq:+part} and \eqref{eq:-part} above.

We continue by exemplifying these relations for a particle near a planar surface.

\section{Example: Lateral force for a nonreciprocal particle near an isotropic plate}
\label{sec:force_example}
We investigate the propulsion force acting on a small nonreciprocal particle placed near a planar surface, the latter made of reciprocal and isotropic material (see Fig.~\ref{fig:setup}). This setting has been calculated in Ref.~\cite{Milton_propulsion_force}. While reproducing the results of Ref.~\cite{Milton_propulsion_force}, we add the possibility for $T_1\not=T_{\mathrm{env}}$~\footnote{In Ref.~\cite{Milton_propulsion_force}, $T_1=T_{\mathrm{env}}$. In this case, the interaction force is canceled out by the environment, and it is sufficient to calculate the self force, see Eq.~\eqref{eq:total_force}.},  we provide new quantitative results for specific materials, and we discuss bounds of the force below. The case of a small reciprocal object in front of a nonreciprocal plate was analyzed in Ref.~\cite{Khandekar_nonreciprocalPlate}.

\subsection{Dielectric plate}
\label{subsec:force_example_dielectric}
We consider a dielectric plate and start with the self force in Eq.~\eqref{eq:F_ss_trace_symmery}. In the point particle limit for object $2$, we use Eq.~\eqref{eq:T2ss_alpha} for the scattering operator $\amsmathbb{T}_2$, and Eq.~\eqref{eq:alpha2} for the polarizability matrix. The Green's function of the plate, $ \amsmathbb{G}_1 $, can be written via the plane waves and Fresnel reflection coefficients $ r_1^N $ and $ r_1^M $ (see, e.g., Eq.~(B21) in Ref.~\cite{Asheichyk_heatradiation}). Using this, the trace in Eq.~\eqref{eq:F_ss_trace_symmery} can be performed, and we obtain
\begin{align}
\notag F_{2,y}^{(2)}(T_2)=&-4\int_0^\infty\! \frac{d\omega}{\omega} \Theta(\omega,T_2)\\
&\times\biggl\{\int_{0}^\infty\frac{ dk_\perp}{2\pi}k_\perp^3\textrm{Im}\left[r^N_1 e^{2ik_zd}\right]\textrm{Im}[\alpha_{2\textrm{f}}]\biggr\}, 
\label{eq:self_force_1order}
\end{align}
where $k_z = \sqrt{\omega^2/c^2-k_\perp^2}$. Only the electric Fresnel coefficient~\cite{TraceFormulas, Asheichyk_heatradiation}
\begin{equation}
r_1^N = \frac{\varepsilon_1\sqrt{\frac{\omega^2}{c^2}-k_{\perp}^2}-\sqrt{\varepsilon_1\frac{\omega^2}{c^2}-k_{\perp}^2}}{\varepsilon_1\sqrt{\frac{\omega^2}{c^2}-k_{\perp}^2}-\sqrt{\varepsilon_1\frac{\omega^2}{c^2}-k_{\perp}^2}},
\label{eq:rN}
\end{equation}
with $ \varepsilon_1 = \varepsilon_1 (\omega) $ being the permittivity of the plate, contributes to the force. Equation~\eqref{eq:self_force_1order} agrees with Eq.~(8.2) in \cite{Milton_propulsion_force}. The force involves $\alpha_{2\textrm{f}}$, showing that it corresponds to the second term in Eq.~\eqref{eq:F_ss_trace_symmery}. As expected, the first term in Eq.~\eqref{eq:F_ss_trace_symmery} does not contribute for a reciprocal plate. The force in Eq.~\eqref{eq:self_force_1order}, as well as the interaction force in Eq.~\eqref{eq:F_y_firstorder}, is proportional to the volume of the particle.

For small distance, the integral over $k_\perp$ in Eq.~\eqref{eq:self_force_1order} can be done to yield
\begin{align}
\notag \lim_{d\ll\lambda_2} F_{2,y}^{(2)}(T_2) = & -\frac{3}{4\pi d^4}\int_0^\infty\! \frac{d\omega}{\omega} \Theta(\omega,T_2)\\
& \times \textrm{Im}\left[\frac{\varepsilon_1-1}{\varepsilon_1+1}\right]\textrm{Im}[\alpha_{2\textrm{f}}].
\label{eq:self_force_1order_nf2}
\end{align}
The force in Eq.~\eqref{eq:self_force_1order_nf2} is proportional to $R^3/d^4$, and it is thus, from these, of similar type compared to perpendicular forces between a nanoparticle and a plate~\cite{Antezza_atom_plate,  Kruger_spheres, TraceFormulas}. The same scaling is also found for the small frequency limit in Eq.~(8.8) in Ref.~\cite{Milton_propulsion_force}. In contrast, the lateral force for an ellipsoid near an isotropic plate is of second order of the ellipsoid's polarizability, yielding a force $\sim R^6/d^7$ \cite{Muller_ellipsoid}; In the considered limit of $R\ll d$, the force $\sim R^3/d^4$ for a nonreciprocal particle is thus expected to be larger than the force $\sim R^6/d^7$ for a reciprocal ellipsoid. 

From Eq.~\eqref{eq:Fps_sym}, the interaction force, in the point particle limit, yields
\begin{align}
\notag F_{1,y}^{(2)}(T_1) = & -4\int_0^\infty\! \frac{d\omega}{\omega} \Theta(\omega,T_1)\\
\notag & \times \bigg\{-\int_{\frac{\omega}{c}}^\infty\frac{ dk_\perp}{2\pi}k_\perp^3e^{-2|k_z|d}
\textrm{Im}\left[r^N_1\right]\textrm{Im}\left[\alpha_{2\textrm{f}}\right]\\
& \ \ \ \ \ \ +\int_0^{\frac{\omega}{c}}\frac{ dk_\perp}{2\pi}k_\perp^3\frac{1}{2}\left(1-|r^N_1|^2\right) \textrm{Im}[\alpha_{2\textrm{s}}]\bigg\}.
\label{eq:F_y_firstorder}
\end{align} 
Equation~\eqref{eq:F_y_firstorder} contains two terms, corresponding to the two terms in Eq.~\eqref{eq:Fps_sym}. The first term involves $\alpha_{2\textrm{f}}$, i.e., the nonreciprocal component. It is a near field term, and it will dominate at small distance. The second term involves  $\alpha_{2\textrm{s}}$, i.e., it gives a force for a reciprocal anisotropic particle. This term is fueled by $(\amsmathbb{G}_1\amsmathbb{V}_{1\textrm{I}}\amsmathbb{G}_1^\dagger)^-$, which, for a reciprocal plate, however, only contributes via propagating modes. For small $d$, this term does not contribute,
\begin{align}
\notag \lim_{ d\ll\lambda_1}\! F^{(2)}_{1,y}(T_1)&=\frac{3}{4\pi d^4}\int_0^\infty\! \frac{d\omega}{\omega} \Theta(\omega,T_1) \textrm{Im}\left[\frac{\varepsilon_1-1}{\varepsilon_1+1}\right]\textrm{Im}[\alpha_{2\textrm{f}}]\\
&=-\lim_{ d\ll\lambda_1}\! F^{(2)}_{2,y}(T_1).
\label{eq:F_y_nonreci_small_d}
\end{align}
The last line emphasizes that  interaction force and self force are equal in magnitude and opposite in sign at small distance.

\begin{figure}[!t]
\begin{center}
\includegraphics[width=0.95\linewidth]{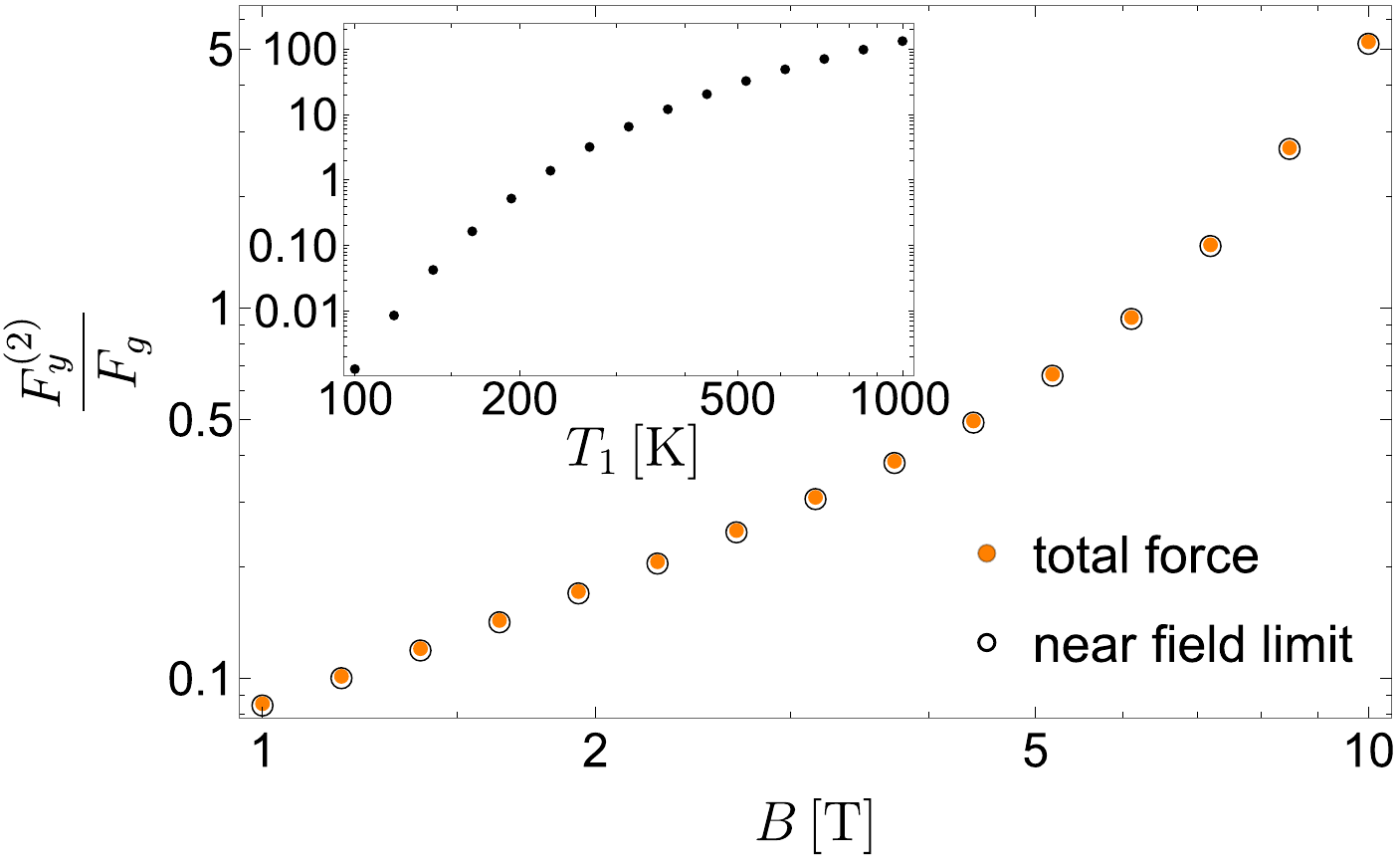}
\end{center}
\caption{\label{fig:force_absorbing_plate}Propulsion force acting on a small sphere modeled by $n$-doped InSb at a distance $d = 100 \ \textrm{nm} $ from a reciprocal isotropic semi-infinite plate [see Fig.~\ref{fig:setup} and Eq.~\eqref{eq:model_plate}], as a function of field strength $B$. The temperatures of the plate, sphere, and environment are $T_1=300 \ \textrm{K}$,  $T_2=10 \ \textrm{K}$, and $T_{\textrm{env}}=0 \ \textrm{K}$, respectively. Orange dots indicate the  force, while the black circles represent the near-field limit obtained from Eqs.~\eqref{eq:self_force_1order_nf2} and~\eqref{eq:F_y_nonreci_small_d}. The force is normalized by the gravitational force $ F_{\textrm{g}} $, with density of the sphere $\rho=5.78\,\mathrm{g}/\mathrm{cm}^3$. Inset depicts the force as a function of  temperature $T_1$ at fixed $B=10\,\mathrm{T}$.}
\end{figure}

The total force is obtained according to Eq.~\eqref{eq:total_force}.
Figure \ref{fig:force_absorbing_plate} shows this force  for a particle made of $n$-doped InSb [see Eqs.~\eqref{eq:eps}] near a dielectric surface, the latter modeled by
\begin{equation}
    \varepsilon_1(\omega)=1+\frac{C_1\omega_1^2}{\omega_1^2-\omega^2-i\gamma_1\omega}.
    \label{eq:model_plate}
\end{equation}
Here, $C_1=2$, $\omega_1=1.15\times 10^{14} \ \textrm{rad} \ \textrm{s}^{-1} $, and $\gamma_1=7\times 10^{10} \ \textrm{rad} \ \textrm{s}^{-1} $ are used, which we found to optimize the overlap with $ \textrm{Im}[\alpha_{2\textrm{f}}] $ in the frequency integral. The figure uses $T_1=300 \ \textrm{K} $, $T_2=10 \ \textrm{K} $, and $T_{\textrm{env}}=0 \ \textrm{K}$.

For small distances $d$ between plate and particle, and especially for  large temperature, the force can be large compared to the gravitational force. The choice of materials is crucial for obtaining a force of significant magnitude.

\subsection{Perfectly conducting plate}
\label{subsec:force_example_pc}
Notably, the second term in Eq.~\eqref{eq:F_ss_trace_symmery} in principle allows a propulsion force for an isolated particle in free space, using $\amsmathbb{G}_0^+=\amsmathbb{G}_0$. However, for a spherical particle with magnetic field pointing in $x$ direction, the force in $y$ direction vanishes by geometric symmetry (for completeness, we mention that there is no force in $x$-direction either). A simple way to break the symmetry is by introducing a mirror, i.e., by evaluating Eq.~\eqref{eq:self_force_1order} in the limit of a perfectly reflecting plate (note that, in this case, there is no interaction force). This yields
\begin{align}
\notag  & \lim_{|\varepsilon_1|\to\infty} F_{2,y}^{(2)}(T_2) = -\frac{1}{4\pi d^4}\int_0^\infty\! \frac{d\omega}{\omega} \Theta(\omega,T_2) \textrm{Im}[\alpha_{2\textrm{f}}]\\
& \times \!\left[3\sin\left(2\frac{\omega}{c}d\right)-6\frac{\omega}{c}d\cos\left(2\frac{\omega}{c}d\right) -4\left(\frac{\omega}{c}d\right)^2\sin\left(2\frac{\omega}{c}d\right)\right].
\label{eq:self_force_1order_pc}
\end{align}
Equation~\eqref{eq:self_force_1order_pc} corresponds to Eq.~(8.11) in Ref.~\cite{Milton_propulsion_force}. Again, we are interested in the distance behaviour. In the far field, the asymptotic form of Eq.~\eqref{eq:self_force_1order_pc} is
\begin{align}
\notag \lim_{d \gg \lambda_2} \lim_{|\varepsilon_1|\to\infty} F_{2,y}^{(2)}(T_2) = & \ \frac{1}{\pi c^2 d^2}\int_0^\infty\! \frac{d\omega}{\omega} \Theta(\omega,T_2)\\
& \times \textrm{Im}[\alpha_{2\textrm{f}}]\sin\left(2\frac{\omega}{c}d\right).
\label{eq:self_force_1order_pc_ff}
\end{align}
This yields an oscillatory behavior, decaying to zero with distance $d$, which reiterates that a sphere in isolation feels no propulsion force. For small distance $d$,
\begin{equation}
\lim_{d\ll\lambda_2} \lim_{|\varepsilon_1|\to\infty}F^{(2)}_{2,y}(T_2)=-\frac{8}{15\pi}d\int_0^\infty\! \frac{d\omega}{\omega} \Theta(\omega,T_2)\textrm{Im}[\alpha_{2\textrm{f}}].
\label{eq:F_ss_conducting_smalld}
\end{equation}
Unlike for plates with finite conductivity, this force does not diverge for small $d$, being, instead, linear in $d$ for small distance. 

\begin{figure}[!t]
\begin{center}
\includegraphics[width=0.95\linewidth]{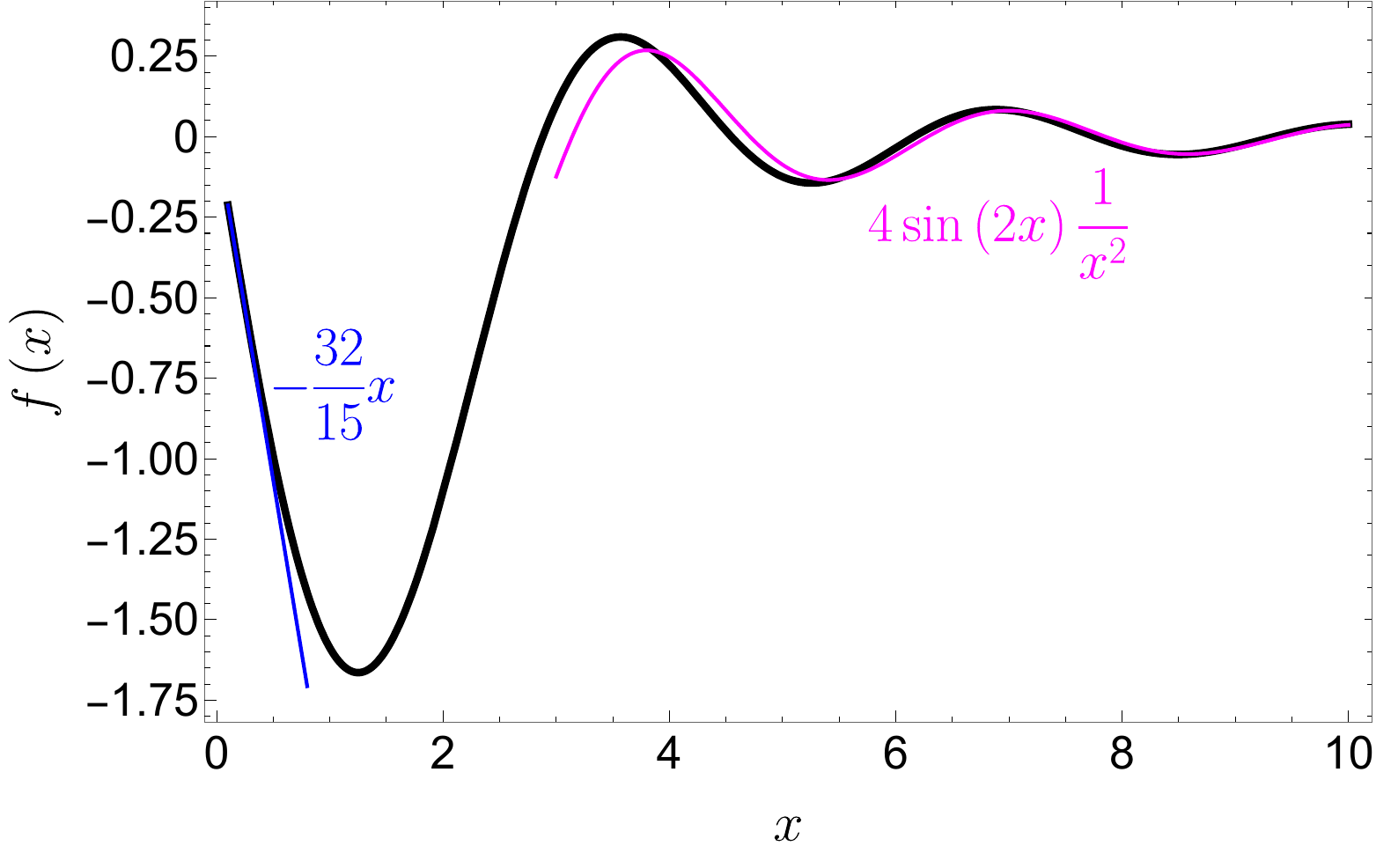}
\end{center}
\caption{\label{fig:force_conducting_plate_2}Function $ f(x) $ defined in Eq.~\eqref{eq:fforce2} and showing the distance dependence of the propulsion force~\eqref{eq:fforce} acting on a nonreciprocal small sphere [modeled by Eq.~\eqref{eq:toy_alpha}] in front of a perfectly reflecting plate.}    
\end{figure}

The force in Eq.~\eqref{eq:self_force_1order_pc} is maximal at a certain distance, which can be found using a simple toy model, 
\begin{equation}
\mathrm{Im}[\alpha_{2\textrm{f}}(\omega)]=\alpha_0\delta(\omega-\omega_0),
\label{eq:toy_alpha}
\end{equation}
with $ \alpha_0 $ and $ \omega_0 $ positive.
For this model, the self force is
\begin{equation}
F_{2,y}^{(2)}(T_2) = \frac{1}{4\pi c^4}\Theta(\omega_0,T_2)\omega_0^3 \alpha_0 f\left(\frac{\omega_0}{c}d\right),
\label{eq:fforce}
\end{equation}
with 
\begin{equation}
f(x) = \frac{1}{x^4}\left[-3\sin(2x)+6x\cos(2x) +4x^2\sin(2x)\right].
\label{eq:fforce2}
\end{equation}
The dimensionless $ f(x) $ as a function of dimensionless $ x = \omega_0d/c $ is plotted in Fig.~\ref{fig:force_conducting_plate_2}. In the near-field limit (small $ x $), $ f(x) = -32x/15 $, while in the far-field limit (large $ x $), $ f(x) = 4\sin(2x)/x^2 $, corresponding to using Eq.~\eqref{eq:toy_alpha} in Eqs.~\eqref{eq:F_ss_conducting_smalld} and~\eqref{eq:self_force_1order_pc_ff}, respectively. The maximal absolute value of $ f (x) $ appears at $ x \approx 5/4 $.

While the sphere experiences a finite force, the $y$ component of the force acting on the perfectly conducting plate is zero. The sphere, in this case, thus exchanges momentum with the environment.

\section{Bound for force in terms of heat transfer}
\label{sec:Bound}
\subsection{General}
\label{subsec:Bound_general}
We show in Appendix~\ref{app:pos} that, if object $ 1 $ is translationally invariant in direction $y$, one has a Fourier decomposition along $y$,
\begin{equation}
\amsmathbb{G}_{1\textrm{I}}(y-y')=\int\frac{dk_y}{2\pi} \hat{\amsmathbb{G}}_{1\textrm{I}},
\label{eq:GFourier}
\end{equation}
with $\hat{\amsmathbb{G}}_{1\textrm{I}}$ nonnegative if  $\amsmathbb{V}_{1\textrm{I}}\geq 0$. We thus write for self emission and self force from Eqs.~\eqref{eq:HT22onescat} and \eqref{eq:F_ss_trace_symmery}, respectively,
\begin{align}
H_2^{(2)} &= \frac{2}{\pi}\int_0^{\infty}\!\! d\omega \Theta(\omega,T_2)\int \frac{dk_y}{2\pi} \Tr\left\{\amsmathbb{T}_{2\textrm{I}}\hat{\amsmathbb{G}}_{1\textrm{I}}\right\},\label{eq:HFourier}\\
F_{2,y}^{(2)}
&=-\frac{2}{\pi }\int_0^\infty\! \frac{d\omega}{\omega} \Theta(\omega,T_2)\int \frac{dk_y}{2\pi} k_y\textrm{Tr}\left\{\amsmathbb{T}_{2\textrm{I}}\hat{\amsmathbb{G}}_{1\textrm{I}}\right\}.\label{eq:FFourier}
\end{align}
As $\textrm{Tr}\left\{\amsmathbb{T}_{2\textrm{I}}\hat{\amsmathbb{G}}_{1\textrm{I}}\right\}\geq 0$, the above yields a bound relation between spectral self force $f_{2,y}^{(2)}(\omega)$ and spectral self emission $h_2^{(2)}(\omega)$ (compare Ref.~\cite{Gelbwaser_nonreciprocalPlate}),
\begin{equation}
\left|f_{2,y}^{(2)}(\omega)\right|c\leq h_2^{(2)}(\omega)\frac{c}{\omega}k_y^{\textrm{max}}.
\label{eq:bound}
\end{equation}
As $k_y^{\textrm{max}}$ is typically set by the inverse distance, this analysis yields a very similar bound between force and energy loss as found in Ref.~\cite{Gelbwaser_nonreciprocalPlate} for two parallel plates.

For heat transfer and interaction force, starting from Eqs.~\eqref{eq:HT12general} (expanded in linear order in $\amsmathbb{T}_2$) and~\eqref{eq:Fps_sym}, the relation analogous to Eq.~\eqref{eq:bound} can be found:
\begin{equation}
\left|f_{1,y}^{(2)}(\omega)\right|c\leq h_1^{(2)}(\omega)\frac{c}{\omega}k_y^{\textrm{max}}
\label{eq:bound2}.
\end{equation}

\subsection{Illustration for a particle near a plate}
\label{subsec:Bound_illustration}
Here we aim to explicitly compare the propulsion force acting on a nonreciprocal particle above a reciprocal isotropic plate to the heat absorbed by the particle. For an illustration, we consider the near-field limit, where the self and interaction parts of the force and heat transfer are equal in absolute value and take simple forms. The self force is given by Eq.~\eqref{eq:self_force_1order_nf2}, while the self emission can be found from Eq.~\eqref{eq:HT22PP_matrix_decompose} (recall that it is independent of $\alpha_{2\textrm{f}}$, as only the $++$ term contributes due to the plate's reciprocity),
\begin{align}
\notag \lim_{d\ll\lambda_{T_2}}H_2^{(2)} = & \ \frac{1}{4\pi d^3}\int_0^\infty\!\! d\omega \Theta(\omega,T_2)\\
& \times \Im\left[\frac{\varepsilon_1-1}{\varepsilon_1+1}\right]\Im[\alpha_{2\textrm{p}}+3\alpha_{2\textrm{d}}],
\label{eq:HT22plateNF}
\end{align}
which can be easily compared with Eq.~\eqref{eq:self_force_1order_nf2}. Indeed, for a passive material, $\mathrm{Im}[\alpha_{2\mathrm{p}}]\geq0$ and $\mathrm{Im}[\alpha_{2\mathrm{d}}]\geq|\mathrm{Im}[\alpha_{2\mathrm{f}}]|$, such that for the spectra of Eqs.~\eqref{eq:self_force_1order_nf2} and~\eqref{eq:HT22plateNF} we have
\begin{equation}
\left|f_{2,y}^{(2)}(\omega)\right|c\leq h_2^{(2)}(\omega)\frac{c}{\omega}\frac{1}{d}
\label{eq:bound3}.
\end{equation}
Since, in the near field, $k^\mathrm{max}_y\approx 1/d$~\cite{Gelbwaser_nonreciprocalPlate}, Eq.~\eqref{eq:bound3} reproduces the general result of Eq.~\eqref{eq:bound} and provides an explicit application.

\section{Conclusion}
\label{sec:concl}
We derive and analyze general formulas for heat transfer and forces involving nonreciprocal nanoparticles. We find that the reciprocal (+) and nonreciprocal parts ($-$) of the particle  couple distinctly to reciprocal (+) and nonreciprocal parts ($-$) of its surroundings. 

The self emission of a nonreciprocal nanoparticle contains ++ and $--$ terms. We  exemplify these for two nonreciprocal nanoparticles, finding that the self emission shows pronounced dependence on the angle between the vector $\vct{d}$ connecting the particles and the pseudovector (e.g., magnetic field) $\vct{B}$ giving rise to nonreciprocity. This is different to the case of reciprocal anisotropic particles studied before~\cite{Incardone_2014, Nikbakht_2014}, where the angular dependence is governed by the particle axis (a director) $\vct{n}$. Despite this, there is apparently no striking difference between reciprocal and nonreciprocal cases for self emission. 

Heat transfer, in contrast, contains a $-+$ term, which yields a so-called ``persistent'' current in a situation with all temperatures equal. While it had been observed before~\cite{Zhu_epsilon_nonreci} for three particles, we here observe it for two nonidentical particles. Notably, it relies on the nonreciprocity of one of the particles and the anisotropy of the other, and occurs only for specific angles between $\vct{d}$, $\vct{n}$, and $\vct{B}$. 

The propulsion force for a nonreciprocal particle in a translationally invariant surrounding is dominated by the $-+$ contribution, so that it is distinct from the force for a reciprocal particle~\cite{Muller_ellipsoid}: It is proportional to the particle volume, while for a reciprocal particle, it is proportional to the volume squared. We find that, for a nonreciprocal particle above a plate~\cite{Milton_propulsion_force}, the force can  be orders of magnitude larger than the gravitational force, so that we expect it to be experimentally detectable. 

Despite the distinct properties of the mentioned $-+$ terms in transfer and force, these are not independent of the $++$ or $--$ terms. The heat transfer, including the $-+$ term, is bound by the $++$ and $--$ term of the self emission. For the propulsion force in a reciprocal surrounding, the $-+$ term is bound by the $++$ term in the heat transfer. This bound  limits the efficiency of this setting as a heat engine~\cite{Gelbwaser_nonreciprocalPlate}, and among other things, it rests on the passivity of the involved objects.

Future work will investigate the consequences of persistent heat current, and also explore ways to break the bound between force and heat transfer. One possibility may be by considering settings without translational invariance such as periodic modulations. Also, particles with magnetic susceptibility or  time or space varying magnetic fields may allow for new phenomena and open new possibilities.

\section*{Acknowledgments}
This joint research project was financially supported by the state of Lower-Saxony and the VolkswagenFoundation, Hannover, Germany (M.K.). K.~A. acknowledges funding by the Deutsche Forschungsgemeinschaft (DFG, German Research Foundation) through the Walter Benjamin fellowship (Project No. 453458207).

\appendix
\section{Free Green's function}
\label{app:G0}
The free Green's function is known in closed form as \cite{Asheichyk_heatradiation}
\begin{align}
\notag \amsmathbb{G}_0(\vct{r},\vct{r'})= & -\frac{1}{3}\frac{c^2}{\omega^2}\mathcal{I}\delta^{(3)}(\vct{r}-\vct{r'})\\
&+\frac{e^{i\frac{\omega}{c}d}}{4\pi d^5}\frac{c^2}{\omega^2}\left[d^2p\mathcal{I}+q(\vct{r}-\vct{r}')\otimes(\vct{r}-\vct{r}')\right],
\label{eq:G0_closed}
\end{align}
where $ p $ and $ q $ are functions of $ \frac{\omega}{c} d $,
\begin{subequations}
\begin{align}
& p \equiv p\left(\frac{\omega}{c}d\right) = -1+i\frac{\omega}{c}d+\frac{\omega^2}{c^2}d^2,\label{eq:p}\\
& q \equiv q\left(\frac{\omega}{c}d\right) = 3-3i\frac{\omega}{c}d-\frac{\omega^2}{c^2}d^2,\label{eq:q}
\end{align}
\label{eq:pq}
\end{subequations}
and $d=|\vct{r}-\vct{r}'|$.

\section{Point particle limit}
\label{app:PP2}
In the point particle limit, we are able to express the scattering operator in terms of the polarizability, which is usually a better known quantity. We assume a spherical nonmagnetic particle small compared to all other length scales of the system, namely all thermal wavelengths $ \lambda_{T_i} = \hbar c/(k_{\textrm{B}}T_i) $, the particle's skin depth, and the distances to other objects. In this point particle limit, the free Green's function inside the particle can be approximated by only the local term [the first term in Eq.~\eqref{eq:G0_closed}]~\cite{Absorption_and_scatteering,Tsang_Mie}, and the scattering operator of the particle simplifies to
\begin{equation}
\amsmathbb{T} = \amsmathbb{V}\left(\amsmathbb{I}+\frac{1}{3}\frac{c^2}{\omega^2}\amsmathbb{V}\right)^{-1}.
\label{eq:T2ss}
\end{equation}
Using $ \amsmathbb{V} = \frac{\omega^2}{c^2}(\bbespilon-\amsmathbb{I}) $ in Eq.~\eqref{eq:T2ss}, we obtain
\begin{equation}
\amsmathbb{T} = 3\frac{\omega^2}{c^2}(\bbespilon-\amsmathbb{I})\left(\bbespilon+2\amsmathbb{I}\right)^{-1}.
\label{eq:T2ssEps}
\end{equation}

For a small particle, $ \bbespilon $ is assumed to be local and homogeneous, i.e., $ \bbespilon(\vct{r},\vct{r}';\omega) = \overset{\leftrightarrow}{\varepsilon}(\omega)\delta^{(3)}(\vct{r}-\vct{r}') $, such that Eq.~\eqref{eq:T2ssEps} can be written as
\begin{equation}
\amsmathbb{T}(\vct{r},\vct{r}') = 3\frac{\omega^2}{c^2}\left(\overset{\leftrightarrow}{\varepsilon}-\mathcal{I}\right)\left(\overset{\leftrightarrow}{\varepsilon}+2\mathcal{I}\right)^{-1}\delta^{(3)}(\vct{r}-\vct{r}').
\label{eq:T2ssMatrix}
\end{equation}
The scattering operator of a small particle is hence local and homogeneous. However, it is in general nondiagonal, as we made no assumptions about $ \overset{\leftrightarrow}{\varepsilon} $. Equation~\eqref{eq:T2ssMatrix} thus holds for anisotropic \cite{Absorption_and_scatteering} as well as magneto-optical nonreciprocal small particles \cite{Lakhtakia_alpha_magneto_optical}. In case of an isotropic and reciprocal particle ($ \overset{\leftrightarrow}{\varepsilon} \propto \mathcal{I} $), Eq.~\eqref{eq:T2ssMatrix} reduces to Eq.~(10) in Ref.~\cite{Asheichyk_heatradiation}. Introducing the polarizability tensor
\begin{equation}
\overset{\leftrightarrow}{\alpha} = \left(\overset{\leftrightarrow}{\varepsilon}-\mathcal{I}\right)\left(\overset{\leftrightarrow}{\varepsilon}+2\mathcal{I}\right)^{-1}R^3,
\label{eq:alpha_def}
\end{equation}
where $ R $ is the particle radius, and the polarizability operator, $ \bbalpha(\vct{r},\vct{r}') = \overset{\leftrightarrow}{\alpha}\delta^{(3)}(\vct{r}-\vct{r}') $, Eq.~\eqref{eq:T2ssMatrix} can be written as
\begin{equation}
\amsmathbb{T}(\vct{r},\vct{r}') = \frac{3}{R^3}\frac{\omega^2}{c^2}\overset{\leftrightarrow}{\alpha}\delta^{(3)}(\vct{r}-\vct{r}') = \frac{3}{R^3}\frac{\omega^2}{c^2}\bbalpha(\vct{r},\vct{r}').
\end{equation}

\section{Permittivity and polarizability tensor of the small particle}
\label{app:permittivityTensor}
The entries of the polarizability tensor Eq.~\eqref{eq:T2ss_alpha} can be determined in terms of the entries of the permittivity tensor Eq.~\eqref{eq:epsilon2} as
\begin{subequations}
\begin{align}
& \alpha_{\textrm{p}} = \frac{\varepsilon_{\textrm{p}}-1}{\varepsilon_{\textrm{p}}+2}R^3,\\
& \alpha_{\textrm{d}} = \left[1 - \frac{3(\varepsilon_{\textrm{d}}+2)}{(\varepsilon_{\textrm{d}}+2)^2-(\varepsilon_{\textrm{s}}^2+\varepsilon_{\textrm{f}}^2)}\right]R^3,\\
& \alpha_{\textrm{s}} = \frac{3\varepsilon_{\textrm{s}}}{(\varepsilon_{\textrm{d}}+2)^2-(\varepsilon_{\textrm{s}}^2+\varepsilon_{\textrm{f}}^2)}R^3,\\
& \alpha_{\textrm{f}} = \frac{3\varepsilon_{\textrm{f}}}{(\varepsilon_{\textrm{d}}+2)^2-(\varepsilon_{\textrm{s}}^2+\varepsilon_{\textrm{f}}^2)}R^3.
\end{align}
\end{subequations}

In numerical computations, we choose the following model for $ \overset{\leftrightarrow}{\varepsilon} $~\cite{Gelbwaser_nonreciprocalPlate, Gelbwaser_nonreciprocalEq, Zhu_epsilon_nonreci, Fan2020}:
\begin{subequations}\label{eq:eps}
\begin{align}
& \varepsilon_{\textrm{p}} = \varepsilon_{\infty} - \frac{\omega_{\textrm{p}}^2}{\omega(\omega+i\omega_{\tau})},\label{eq:ep}\\
& \varepsilon_{\textrm{d}} = \varepsilon_{\infty} - \frac{\omega_{\textrm{p}}^2(\omega+i\omega_{\tau})}{\omega[(\omega+i\omega_{\tau})^2-\omega_B^2]},\label{eq:ed}\\
& \varepsilon_{\textrm{f}} = -\frac{\omega_B\omega_{\textrm{p}}^2}{\omega[(\omega+i\omega_{\tau})^2-\omega_B^2]},\label{eq:ef}\\
& \varepsilon_{\textrm{s}} = 0,\label{eq:es}
\end{align}
\end{subequations}
where $ \omega_B $ is the cyclotron frequency due to the external magnetic field breaking reciprocity; $ \omega_{\textrm{p}} $ and $ \omega_{\tau} $ are the plasma and damping frequency, respectively, and $ \varepsilon_{\infty} $ is the high-frequency dielectric constant. For the material of the particle we choose $n$-doped InSb, for which $ \omega_{\textrm{p}} = 7.4 \times 10^{14} \ \textrm{rad} \ \textrm{s}^{-1} $, $ \omega_{\tau} = 6.3 \times 10^{12} \ \textrm{rad} \ \textrm{s}^{-1} $, and $ \varepsilon_{\infty} = 15.7 $~\cite{Zhu_epsilon_nonreci}. The value of $ \omega_B = 2.2 \times 10^{12} \times B \ \textrm{rad} \ \textrm{s}^{-1} $ is varied throughout computations via the variation of the magnetic field strength $ B $. For any $ B $, the skin depth corresponding to any direction (estimated as $ 1/(k\Im\sqrt{\varepsilon_{\{\textrm{p},\textrm{d},\textrm{f}\}}} $), is found to be larger than $ 400 \ \textrm{nm} $, such that we can use a particle with $ R \approx 50 \ \textrm{nm} $ or smaller for the point particle limit to be valid.

\section{Traces for the self emission of a point particle}
\label{app:traces_self_emission}
Splitting the Green's function of particle $1$ and the polarizability of particle $2$ in their symmetric and antisymmetric parts, we compute the two traces in Eq.~\eqref{eq:HT22PP_matrix_decompose}:
\begin{widetext}
\begin{subequations}
\begin{align}
\notag & 4\pi\frac{\omega^2}{c^2}d^{10}\Tr\left\{\overset{\leftrightarrow}{\alpha}_{2\textrm{I}}^+\amsmathbb{G}_{1\textrm{I}}^+(\vct{r}_2,\vct{r}_2)\right\} = 4\pi\frac{\omega^2}{c^2}d^{10}\frac{1}{6\pi}\frac{\omega}{c}\left(\Im[\alpha_{2\textrm{p}}]+2\Im[\alpha_{2\textrm{d}}]\right)\\
\notag & + \Im[\alpha_{2\textrm{p}}]\Im\left\{e^{2i\frac{\omega}{c}d}\left[\alpha_{1\textrm{p}}\left(pd^2+qx^2\right)^2+\left(\alpha_{1\textrm{d}}+\alpha_{1\textrm{s}}\sin(2\varphi)\right)q^2x^2r^2\right]\right\}\\
\notag & + \Im[\alpha_{2\textrm{d}}]\Im\left\{e^{2i\frac{\omega}{c}d}\left[\alpha_{1\textrm{p}}q^2x^2r^2+\alpha_{1\textrm{d}}\left(2p^2d^4+2pqd^2r^2+q^2r^4\right)+\alpha_{1\textrm{s}}\left(pd^2+qr^2\right)qr^2\sin(2\varphi)\right]\right\}\\
& + \Im[\alpha_{2\textrm{s}}]\Im\left\{e^{2i\frac{\omega}{c}d}\left[\alpha_{1\textrm{p}}q^2x^2r^2\sin(2\varphi)+\alpha_{1\textrm{d}}\left(pqd^2+q^2r^2\sin(2\varphi)\right)r^2+\alpha_{1\textrm{s}}\left(2p^2d^4+2pqd^2r^2+q^2r^4\sin^2(2\varphi)\right)\right]\right\},\label{eq:TrSE++}\\
& \Tr\left\{\overset{\leftrightarrow}{\alpha}_{2\textrm{I}}^-\amsmathbb{G}_{1\textrm{I}}^-(\vct{r}_2,\vct{r}_2)\right\} = \frac{1}{2\pi \frac{\omega^2}{c^2}d^8}\Im[\alpha_{2\textrm{f}}]\Im\left\{\alpha_{1\textrm{f}}e^{2i\frac{\omega}{c}d}p\left(pd^2+qr^2\right)\right\},\label{eq:TrSE--}
\end{align}
\label{eq:TrSE}
\end{subequations}
\end{widetext}
where $ r^2 $ and $ x^2 $ are the squared interparticle distances perpendicular to and along the $x$ axis (the direction of the magnetic field), respectively, and $ \varphi $ is the azimuthal angle in the $ yz $ plane (see Fig.~\ref{fig:setup_2spheres}); $ d = \sqrt{r^2+x^2} $ is the full distance. $ p $ and $ q $ are given by Eq.~\eqref{eq:pq}.
The first term in Eq.~\eqref{eq:TrSE++} corresponds to the self emission in vacuum, i.e., without particle $ 1 $

\section{Positivity of Fourier transforms}
\label{app:pos}
Consider an operator $\amsmathbb{A}=\amsmathbb{A}(y-y')$, which is Hermitian and nonnegative, and translationally invariant in direction $y$. Due to translational invariance, its Fourier transform along $y$, $\tilde{\amsmathbb{A}}(k_y)$, is a function of one wavenumber $k_y$ only,
\begin{equation}
\amsmathbb{A}(y-y')=\int\frac{dk_y}{2\pi}\tilde{\amsmathbb{A}}(k_y)e^{ik_y(y-y')}.
\label{eq:AFourier}
\end{equation}
The inverse of this relation is
\begin{align}
\tilde{\amsmathbb{A}}(k_y)&=\int d(y-y')\amsmathbb{A}(y-y')e^{-ik_y(y-y')}\notag\\
&=\frac{1}{L}\int dydy' e^{-ik_yy} \amsmathbb{A}e^{ik_yy'}=\frac{1}{L} \amsmathbb{F}\amsmathbb{A}\amsmathbb{F}^\dagger.
\label{eq:pos}
\end{align}
We defined $\amsmathbb{F}=\amsmathbb{I}e^{-ik_yy}$, and introduced the formally infinite length $L$ of the system along $y$. Eq.~\eqref{eq:pos} shows that $\tilde{\amsmathbb{A}}$ is nonnegative if $\amsmathbb{A}$ is. The same holds for 
\begin{equation}
\hat{\amsmathbb{A}}= e^{ik_yy} \tilde{ \amsmathbb{A}}e^{-ik_yy'}.
\label{eq:Ahat}
\end{equation}

\section{Tensorial structure for small $B$}
\label{app:tensorial}
In this section, we will analyze the tensorial structure of the expressions in the main text. For small values of $B$, the polarizability tensors take the form (the Einstein summation convention is assumed)
\begin{equation}
\alpha_{ij}=\alpha^{(\textrm{d})}\delta_{ij}+\alpha^{(\textrm{s})}n_i n_j+\alpha^{(\textrm{a})}\epsilon_{ijk} B_k,
\label{eq:alpha_ij}
\end{equation}
or in shorter form,
\begin{equation}
\at=\alpha^{(\textrm{d})}\mathcal{I}+\alpha^{(\textrm{s})}\nt+\alpha^{(\textrm{a})}\Bt,
\label{eq:alpha_tensorial}
\end{equation}
where $ n_i $ is the $i$th component of the optical axis $ \vct{n} $, giving rise to the symmetric off-diagonal elements, and $ B_k $ is the $ k $th component of the magnetic field vector $ \vct{B} $, responsible for the antisymmetric part.
  
The free Green's function [see Eq.~\eqref{eq:G0_closed}] takes the form
\begin{equation}
G_{0,ij}=G_0^{(\textrm{d})}\delta_{ij}+G_0^{(\textrm{s})} d_i d_j,
\label{eq:G_ij}
\end{equation}
or
\begin{equation}
\Gt=G_0^{(\textrm{d})}\mathcal{I}+G_0^{(\textrm{s})} \dt,
\label{eq:G_tensorial}
\end{equation}
with with $d_i$ being the $i$th component of the vector connecting the two arguments of the Green's function. 

For the self emission of a particle near a second one, we have, from Eqs.~\eqref{eq:HT22PP_matrix} and \eqref{eq:G1PP}, terms of the form $\Tr\Big(\at_1\Gt\at_2\Gt\Big)$. We are interested in the dependence on the magnetic field. We analyze only the $--$ term. In this term, the summands $\sim \epsilon_{ijk} B_k$ from both particles must contribute. We thus have the terms
\begin{subequations}\label{eq:BtensorialSE}
\begin{align}
\Tr\left[\Bt\dt\Bt\dt\right]&=\epsilon_{ijk} B_k  d_j d_l \epsilon_{lmr} B_rd_m d_i  =0,\label{eq:BdBd}\\
\notag \Tr\left[\Bt\dt\Bt\mathcal{I}\right]&=\epsilon_{ijk} B_k  d_j d_l \epsilon_{lmr} B_r \delta_{mi}\\
&= -(\vct{d}\times\vct{B})\cdot (\vct{d}\times\vct{B}),\label{eq:BdBI}\\
\Tr\left[\Bt\mathcal{I}\Bt\mathcal{I}\right]&=\epsilon_{ijk} B_k  \delta_{jl} \epsilon_{lmr} B_r \delta_{mi}  = -2 B^2.\label{eq:BIBI}
\end{align}
\end{subequations}
The self emission in presence of a second particle thus carries an isotropic term $\sim B^2$, and a term $\sim(\vct{d}\times\vct{B})\cdot (\vct{d}\times\vct{B})$, corresponding to the first and second term in Eq.~\eqref{eq:TrSE--}, respectively. 

For the persistent current, accoring to Eq.~\eqref{eq:HTnet1PP2PP_antisymm}, the $ +- $ terms, i.e., the symmetric part of one particle and the antisymmetric part of the other one, contribute. Therefore, we collect the terms
\begin{subequations}
\begin{align}
\Tr\left[\nt\dt\Bt\dt\right]&=n_in_j  d_j d_l \epsilon_{lmr} B_rd_m d_i  = 0,\label{eq:ndBd}\\
\Tr\left[\nt\dt\Bt\mathcal{I}\right]&=n_in_j  d_j d_l \epsilon_{lmr} B_r \delta_{mi} = (\vct{n}\cdot \vct{d}) (\vct{n}\times\vct{B})\cdot \vct{d},\label{eq:ndBI}\\
\Tr\left[\nt\mathcal{I}\Bt\mathcal{I}\right]&= n_in_j  \delta_{jl} \epsilon_{lmr} B_r \delta_{mi} = 0.\label{eq:nIBI}
\end{align}
\end{subequations}
Equation~\eqref{eq:ndBI} shows that the persistent current is of form  $ (\vct{n}\cdot \vct{d}) (\vct{n}\times\vct{B})\cdot \vct{d} $. It thus vanishes if $\vct{n}$ is perpendicular to $\vct{d}$, or if any pair of $\vct{n}$, $\vct{B}$, and $\vct{d}$ is parallel, in agreement with Eq.~\eqref{eq:HTnet1PP2PP_explicit}.


\begin{thebibliography}{62}%
\makeatletter
\providecommand \@ifxundefined [1]{%
 \@ifx{#1\undefined}
}%
\providecommand \@ifnum [1]{%
 \ifnum #1\expandafter \@firstoftwo
 \else \expandafter \@secondoftwo
 \fi
}%
\providecommand \@ifx [1]{%
 \ifx #1\expandafter \@firstoftwo
 \else \expandafter \@secondoftwo
 \fi
}%
\providecommand \natexlab [1]{#1}%
\providecommand \enquote  [1]{``#1''}%
\providecommand \bibnamefont  [1]{#1}%
\providecommand \bibfnamefont [1]{#1}%
\providecommand \citenamefont [1]{#1}%
\providecommand \href@noop [0]{\@secondoftwo}%
\providecommand \href [0]{\begingroup \@sanitize@url \@href}%
\providecommand \@href[1]{\@@startlink{#1}\@@href}%
\providecommand \@@href[1]{\endgroup#1\@@endlink}%
\providecommand \@sanitize@url [0]{\catcode `\\12\catcode `\$12\catcode
  `\&12\catcode `\#12\catcode `\^12\catcode `\_12\catcode `\%12\relax}%
\providecommand \@@startlink[1]{}%
\providecommand \@@endlink[0]{}%
\providecommand \url  [0]{\begingroup\@sanitize@url \@url }%
\providecommand \@url [1]{\endgroup\@href {#1}{\urlprefix }}%
\providecommand \urlprefix  [0]{URL }%
\providecommand \Eprint [0]{\href }%
\providecommand \doibase [0]{https://doi.org/}%
\providecommand \selectlanguage [0]{\@gobble}%
\providecommand \bibinfo  [0]{\@secondoftwo}%
\providecommand \bibfield  [0]{\@secondoftwo}%
\providecommand \translation [1]{[#1]}%
\providecommand \BibitemOpen [0]{}%
\providecommand \bibitemStop [0]{}%
\providecommand \bibitemNoStop [0]{.\EOS\space}%
\providecommand \EOS [0]{\spacefactor3000\relax}%
\providecommand \BibitemShut  [1]{\csname bibitem#1\endcsname}%
\let\auto@bib@innerbib\@empty
\bibitem [{\citenamefont {Casimir}(1948)}]{Casimir_original}%
  \BibitemOpen
  \bibfield  {author} {\bibinfo {author} {\bibfnamefont {H.~B.~G.}\
  \bibnamefont {Casimir}},\ }\bibfield  {title} {\bibinfo {title} {On the
  attraction between two perfectly conducting plates},\ }\href@noop {}
  {\bibfield  {journal} {\bibinfo  {journal} {Indag. Math.}\ }\textbf {\bibinfo
  {volume} {10}},\ \bibinfo {pages} {261} (\bibinfo {year} {1948})}\BibitemShut
  {NoStop}%
\bibitem [{\citenamefont {Derjaguin}\ \emph {et~al.}(1956)\citenamefont
  {Derjaguin}, \citenamefont {Abrikosova},\ and\ \citenamefont
  {Lifshitz}}]{Lif_measurement}%
  \BibitemOpen
  \bibfield  {author} {\bibinfo {author} {\bibfnamefont {B.~V.}\ \bibnamefont
  {Derjaguin}}, \bibinfo {author} {\bibfnamefont {I.~I.}\ \bibnamefont
  {Abrikosova}},\ and\ \bibinfo {author} {\bibfnamefont {E.~M.}\ \bibnamefont
  {Lifshitz}},\ }\bibfield  {title} {\bibinfo {title} {Direct measurement of
  molecular attraction between solids separated by a narrow gap},\ }\href
  {https://doi.org/10.1039/QR9561000295} {\bibfield  {journal} {\bibinfo
  {journal} {Q. Rev. Chem. Soc.}\ }\textbf {\bibinfo {volume} {10}},\ \bibinfo
  {pages} {295} (\bibinfo {year} {1956})}\BibitemShut {NoStop}%
\bibitem [{\citenamefont {Lifshitz}(1956)}]{Lif_temp}%
  \BibitemOpen
  \bibfield  {author} {\bibinfo {author} {\bibfnamefont {E.~M.}\ \bibnamefont
  {Lifshitz}},\ }\bibfield  {title} {\bibinfo {title} {The theory of molecular
  attractive forces between solids},\ }\href@noop {} {\bibfield  {journal}
  {\bibinfo  {journal} {Sov. Phys. JETP}\ }\textbf {\bibinfo {volume} {2}},\
  \bibinfo {pages} {73} (\bibinfo {year} {1956})}\BibitemShut {NoStop}%
\bibitem [{\citenamefont {Emig}\ \emph {et~al.}(2007)\citenamefont {Emig},
  \citenamefont {Graham}, \citenamefont {Jaffe},\ and\ \citenamefont
  {Kardar}}]{Emig_scattering}%
  \BibitemOpen
  \bibfield  {author} {\bibinfo {author} {\bibfnamefont {T.}~\bibnamefont
  {Emig}}, \bibinfo {author} {\bibfnamefont {N.}~\bibnamefont {Graham}},
  \bibinfo {author} {\bibfnamefont {R.~L.}\ \bibnamefont {Jaffe}},\ and\
  \bibinfo {author} {\bibfnamefont {M.}~\bibnamefont {Kardar}},\ }\bibfield
  {title} {\bibinfo {title} {Casimir forces between arbitrary compact
  objects},\ }\href {https://doi.org/10.1103/PhysRevLett.99.170403} {\bibfield
  {journal} {\bibinfo  {journal} {Phys. Rev. Lett.}\ }\textbf {\bibinfo
  {volume} {99}},\ \bibinfo {pages} {170403} (\bibinfo {year}
  {2007})}\BibitemShut {NoStop}%
\bibitem [{\citenamefont {Rahi}\ \emph {et~al.}(2009)\citenamefont {Rahi},
  \citenamefont {Emig}, \citenamefont {Graham}, \citenamefont {Jaffe},\ and\
  \citenamefont {Kardar}}]{Rahi_expansionWaves_equilibrium}%
  \BibitemOpen
  \bibfield  {author} {\bibinfo {author} {\bibfnamefont {S.~J.}\ \bibnamefont
  {Rahi}}, \bibinfo {author} {\bibfnamefont {T.}~\bibnamefont {Emig}}, \bibinfo
  {author} {\bibfnamefont {N.}~\bibnamefont {Graham}}, \bibinfo {author}
  {\bibfnamefont {R.~L.}\ \bibnamefont {Jaffe}},\ and\ \bibinfo {author}
  {\bibfnamefont {M.}~\bibnamefont {Kardar}},\ }\bibfield  {title} {\bibinfo
  {title} {Scattering theory approach to electrodynamic {C}asimir forces},\
  }\href {https://doi.org/10.1103/PhysRevD.80.085021} {\bibfield  {journal}
  {\bibinfo  {journal} {Phys. Rev. D}\ }\textbf {\bibinfo {volume} {80}},\
  \bibinfo {pages} {085021} (\bibinfo {year} {2009})}\BibitemShut {NoStop}%
\bibitem [{\citenamefont {Eckhardt}(1984)}]{Eckhardt_correlator_eq_gf}%
  \BibitemOpen
  \bibfield  {author} {\bibinfo {author} {\bibfnamefont {W.}~\bibnamefont
  {Eckhardt}},\ }\bibfield  {title} {\bibinfo {title} {Macroscopic theory of
  electromagnetic fluctuations and stationary radiative heat transfer},\ }\href
  {https://doi.org/10.1103/PhysRevA.29.1991} {\bibfield  {journal} {\bibinfo
  {journal} {Phys. Rev. A}\ }\textbf {\bibinfo {volume} {29}},\ \bibinfo
  {pages} {1991} (\bibinfo {year} {1984})}\BibitemShut {NoStop}%
\bibitem [{\citenamefont {Maia~Neto}\ \emph {et~al.}(2008)\citenamefont
  {Maia~Neto}, \citenamefont {Lambrecht},\ and\ \citenamefont
  {Reynaud}}]{Maia_CasimirEnergy}%
  \BibitemOpen
  \bibfield  {author} {\bibinfo {author} {\bibfnamefont {P.~A.}\ \bibnamefont
  {Maia~Neto}}, \bibinfo {author} {\bibfnamefont {A.}~\bibnamefont
  {Lambrecht}},\ and\ \bibinfo {author} {\bibfnamefont {S.}~\bibnamefont
  {Reynaud}},\ }\bibfield  {title} {\bibinfo {title} {Casimir energy between a
  plane and a sphere in electromagnetic vacuum},\ }\href
  {https://doi.org/10.1103/PhysRevA.78.012115} {\bibfield  {journal} {\bibinfo
  {journal} {Phys. Rev. A}\ }\textbf {\bibinfo {volume} {78}},\ \bibinfo
  {pages} {012115} (\bibinfo {year} {2008})}\BibitemShut {NoStop}%
\bibitem [{\citenamefont {Reid}\ \emph {et~al.}(2011)\citenamefont {Reid},
  \citenamefont {White},\ and\ \citenamefont {Johnson}}]{scatter_conv_eq}%
  \BibitemOpen
  \bibfield  {author} {\bibinfo {author} {\bibfnamefont {M.~T.~H.}\
  \bibnamefont {Reid}}, \bibinfo {author} {\bibfnamefont {J.}~\bibnamefont
  {White}},\ and\ \bibinfo {author} {\bibfnamefont {S.~G.}\ \bibnamefont
  {Johnson}},\ }\bibfield  {title} {\bibinfo {title} {Computation of {C}asimir
  interactions between arbitrary three-dimensional objects with arbitrary
  material properties},\ }\href {https://doi.org/10.1103/PhysRevA.84.010503}
  {\bibfield  {journal} {\bibinfo  {journal} {Phys. Rev. A}\ }\textbf {\bibinfo
  {volume} {84}},\ \bibinfo {pages} {010503} (\bibinfo {year}
  {2011})}\BibitemShut {NoStop}%
\bibitem [{\citenamefont {Lamoreaux}(1997)}]{exp1}%
  \BibitemOpen
  \bibfield  {author} {\bibinfo {author} {\bibfnamefont {S.~K.}\ \bibnamefont
  {Lamoreaux}},\ }\bibfield  {title} {\bibinfo {title} {Demonstration of the
  {C}asimir force in the 0.6 to $6\ensuremath{\mu}m$ range},\ }\href
  {https://doi.org/10.1103/PhysRevLett.78.5} {\bibfield  {journal} {\bibinfo
  {journal} {Phys. Rev. Lett.}\ }\textbf {\bibinfo {volume} {78}},\ \bibinfo
  {pages} {5} (\bibinfo {year} {1997})}\BibitemShut {NoStop}%
\bibitem [{\citenamefont {Garrett}\ \emph {et~al.}(2018)\citenamefont
  {Garrett}, \citenamefont {Somers},\ and\ \citenamefont {Munday}}]{exp2}%
  \BibitemOpen
  \bibfield  {author} {\bibinfo {author} {\bibfnamefont {J.~L.}\ \bibnamefont
  {Garrett}}, \bibinfo {author} {\bibfnamefont {D.~A.~T.}\ \bibnamefont
  {Somers}},\ and\ \bibinfo {author} {\bibfnamefont {J.~N.}\ \bibnamefont
  {Munday}},\ }\bibfield  {title} {\bibinfo {title} {Measurement of the
  {C}asimir force between two spheres},\ }\href
  {https://doi.org/10.1103/PhysRevLett.120.040401} {\bibfield  {journal}
  {\bibinfo  {journal} {Phys. Rev. Lett.}\ }\textbf {\bibinfo {volume} {120}},\
  \bibinfo {pages} {040401} (\bibinfo {year} {2018})}\BibitemShut {NoStop}%
\bibitem [{\citenamefont {Bressi}\ \emph {et~al.}(2002)\citenamefont {Bressi},
  \citenamefont {Carugno}, \citenamefont {Onofrio},\ and\ \citenamefont
  {Ruoso}}]{exp3}%
  \BibitemOpen
  \bibfield  {author} {\bibinfo {author} {\bibfnamefont {G.}~\bibnamefont
  {Bressi}}, \bibinfo {author} {\bibfnamefont {G.}~\bibnamefont {Carugno}},
  \bibinfo {author} {\bibfnamefont {R.}~\bibnamefont {Onofrio}},\ and\ \bibinfo
  {author} {\bibfnamefont {G.}~\bibnamefont {Ruoso}},\ }\bibfield  {title}
  {\bibinfo {title} {Measurement of the {C}asimir force between parallel
  metallic surfaces},\ }\href {https://doi.org/10.1103/PhysRevLett.88.041804}
  {\bibfield  {journal} {\bibinfo  {journal} {Phys. Rev. Lett.}\ }\textbf
  {\bibinfo {volume} {88}},\ \bibinfo {pages} {041804} (\bibinfo {year}
  {2002})}\BibitemShut {NoStop}%
\bibitem [{\citenamefont {Rytov}(1989)}]{Rytov_book}%
  \BibitemOpen
  \bibfield  {author} {\bibinfo {author} {\bibfnamefont {S.~M.}\ \bibnamefont
  {Rytov}},\ }\href@noop {} {\emph {\bibinfo {title} {Principles of Statistical
  Radiophysics 3}}}\ (\bibinfo  {publisher} {Springer-Verlag},\ \bibinfo {year}
  {1989})\BibitemShut {NoStop}%
\bibitem [{\citenamefont {Rytov}(1958)}]{Rytov_paper}%
  \BibitemOpen
  \bibfield  {author} {\bibinfo {author} {\bibfnamefont {S.}~\bibnamefont
  {Rytov}},\ }\bibfield  {title} {\bibinfo {title} {Correlation theory of
  thermal fluctuations in an isotropic medium},\ }\href@noop {} {\bibfield
  {journal} {\bibinfo  {journal} {Sov. Phys. JETP}\ }\textbf {\bibinfo {volume}
  {6}},\ \bibinfo {pages} {130} (\bibinfo {year} {1958})}\BibitemShut {NoStop}%
\bibitem [{\citenamefont {Antezza}\ \emph {et~al.}(2005)\citenamefont
  {Antezza}, \citenamefont {Pitaevskii},\ and\ \citenamefont
  {Stringari}}]{Antezza_atom_plate}%
  \BibitemOpen
  \bibfield  {author} {\bibinfo {author} {\bibfnamefont {M.}~\bibnamefont
  {Antezza}}, \bibinfo {author} {\bibfnamefont {L.~P.}\ \bibnamefont
  {Pitaevskii}},\ and\ \bibinfo {author} {\bibfnamefont {S.}~\bibnamefont
  {Stringari}},\ }\bibfield  {title} {\bibinfo {title} {New asymptotic behavior
  of the surface-atom force out of thermal equilibrium},\ }\href
  {https://doi.org/10.1103/PhysRevLett.95.113202} {\bibfield  {journal}
  {\bibinfo  {journal} {Phys. Rev. Lett.}\ }\textbf {\bibinfo {volume} {95}},\
  \bibinfo {pages} {113202} (\bibinfo {year} {2005})}\BibitemShut {NoStop}%
\bibitem [{\citenamefont {Messina}\ and\ \citenamefont
  {Antezza}(2011)}]{Messina2011}%
  \BibitemOpen
  \bibfield  {author} {\bibinfo {author} {\bibfnamefont {R.}~\bibnamefont
  {Messina}}\ and\ \bibinfo {author} {\bibfnamefont {M.}~\bibnamefont
  {Antezza}},\ }\bibfield  {title} {\bibinfo {title} {Scattering-matrix
  approach to {C}asimir-{L}ifshitz force and heat transfer out of thermal
  equilibrium between arbitrary bodies},\ }\href
  {https://doi.org/10.1103/PhysRevA.84.042102} {\bibfield  {journal} {\bibinfo
  {journal} {Phys. Rev. A}\ }\textbf {\bibinfo {volume} {84}},\ \bibinfo
  {pages} {042102} (\bibinfo {year} {2011})}\BibitemShut {NoStop}%
\bibitem [{\citenamefont {Kr\"uger}\ \emph {et~al.}(2011)\citenamefont
  {Kr\"uger}, \citenamefont {Emig},\ and\ \citenamefont
  {Kardar}}]{Kruger_correlator}%
  \BibitemOpen
  \bibfield  {author} {\bibinfo {author} {\bibfnamefont {M.}~\bibnamefont
  {Kr\"uger}}, \bibinfo {author} {\bibfnamefont {T.}~\bibnamefont {Emig}},\
  and\ \bibinfo {author} {\bibfnamefont {M.}~\bibnamefont {Kardar}},\
  }\bibfield  {title} {\bibinfo {title} {Nonequilibrium electromagnetic
  fluctuations: Heat transfer and interactions},\ }\href
  {https://doi.org/10.1103/PhysRevLett.106.210404} {\bibfield  {journal}
  {\bibinfo  {journal} {Phys. Rev. Lett.}\ }\textbf {\bibinfo {volume} {106}},\
  \bibinfo {pages} {210404} (\bibinfo {year} {2011})}\BibitemShut {NoStop}%
\bibitem [{\citenamefont {Kr\"uger}\ \emph {et~al.}(2012)\citenamefont
  {Kr\"uger}, \citenamefont {Bimonte}, \citenamefont {Emig},\ and\
  \citenamefont {Kardar}}]{TraceFormulas}%
  \BibitemOpen
  \bibfield  {author} {\bibinfo {author} {\bibfnamefont {M.}~\bibnamefont
  {Kr\"uger}}, \bibinfo {author} {\bibfnamefont {G.}~\bibnamefont {Bimonte}},
  \bibinfo {author} {\bibfnamefont {T.}~\bibnamefont {Emig}},\ and\ \bibinfo
  {author} {\bibfnamefont {M.}~\bibnamefont {Kardar}},\ }\bibfield  {title}
  {\bibinfo {title} {Trace formulas for nonequilibrium {C}asimir interactions,
  heat radiation, and heat transfer for arbitrary objects},\ }\href
  {https://doi.org/10.1103/PhysRevB.86.115423} {\bibfield  {journal} {\bibinfo
  {journal} {Phys. Rev. B}\ }\textbf {\bibinfo {volume} {86}},\ \bibinfo
  {pages} {115423} (\bibinfo {year} {2012})}\BibitemShut {NoStop}%
\bibitem [{\citenamefont {Rodriguez}\ \emph
  {et~al.}(2012{\natexlab{a}})\citenamefont {Rodriguez}, \citenamefont {Reid},\
  and\ \citenamefont {Johnson}}]{Rodriguez_trace}%
  \BibitemOpen
  \bibfield  {author} {\bibinfo {author} {\bibfnamefont {A.~W.}\ \bibnamefont
  {Rodriguez}}, \bibinfo {author} {\bibfnamefont {M.~T.~H.}\ \bibnamefont
  {Reid}},\ and\ \bibinfo {author} {\bibfnamefont {S.~G.}\ \bibnamefont
  {Johnson}},\ }\bibfield  {title} {\bibinfo {title}
  {Fluctuating-surface-current formulation of radiative heat transfer for
  arbitrary geometries},\ }\href {https://doi.org/10.1103/PhysRevB.86.220302}
  {\bibfield  {journal} {\bibinfo  {journal} {Phys. Rev. B}\ }\textbf {\bibinfo
  {volume} {86}},\ \bibinfo {pages} {220302} (\bibinfo {year}
  {2012}{\natexlab{a}})}\BibitemShut {NoStop}%
\bibitem [{\citenamefont {Polder}\ and\ \citenamefont
  {Van~Hove}(1971)}]{Polder_HT_plates}%
  \BibitemOpen
  \bibfield  {author} {\bibinfo {author} {\bibfnamefont {D.}~\bibnamefont
  {Polder}}\ and\ \bibinfo {author} {\bibfnamefont {M.}~\bibnamefont
  {Van~Hove}},\ }\bibfield  {title} {\bibinfo {title} {Theory of radiative heat
  transfer between closely spaced bodies},\ }\href
  {https://doi.org/10.1103/PhysRevB.4.3303} {\bibfield  {journal} {\bibinfo
  {journal} {Phys. Rev. B}\ }\textbf {\bibinfo {volume} {4}},\ \bibinfo {pages}
  {3303} (\bibinfo {year} {1971})}\BibitemShut {NoStop}%
\bibitem [{\citenamefont {Narayanaswamy}\ and\ \citenamefont
  {Chen}(2008{\natexlab{a}})}]{Narayanaswamy_HT_spheres}%
  \BibitemOpen
  \bibfield  {author} {\bibinfo {author} {\bibfnamefont {A.}~\bibnamefont
  {Narayanaswamy}}\ and\ \bibinfo {author} {\bibfnamefont {G.}~\bibnamefont
  {Chen}},\ }\bibfield  {title} {\bibinfo {title} {Thermal near-field radiative
  transfer between two spheres},\ }\href
  {https://doi.org/10.1103/PhysRevB.77.075125} {\bibfield  {journal} {\bibinfo
  {journal} {Phys. Rev. B}\ }\textbf {\bibinfo {volume} {77}},\ \bibinfo
  {pages} {075125} (\bibinfo {year} {2008}{\natexlab{a}})}\BibitemShut
  {NoStop}%
\bibitem [{\citenamefont {Antezza}\ \emph {et~al.}(2008)\citenamefont
  {Antezza}, \citenamefont {Pitaevskii}, \citenamefont {Stringari},\ and\
  \citenamefont {Svetovoy}}]{Antezza08}%
  \BibitemOpen
  \bibfield  {author} {\bibinfo {author} {\bibfnamefont {M.}~\bibnamefont
  {Antezza}}, \bibinfo {author} {\bibfnamefont {L.~P.}\ \bibnamefont
  {Pitaevskii}}, \bibinfo {author} {\bibfnamefont {S.}~\bibnamefont
  {Stringari}},\ and\ \bibinfo {author} {\bibfnamefont {V.~B.}\ \bibnamefont
  {Svetovoy}},\ }\bibfield  {title} {\bibinfo {title} {Casimir-{L}ifshitz force
  out of thermal equilibrium},\ }\href
  {https://doi.org/10.1103/PhysRevA.77.022901} {\bibfield  {journal} {\bibinfo
  {journal} {Phys. Rev. A}\ }\textbf {\bibinfo {volume} {77}},\ \bibinfo
  {pages} {022901} (\bibinfo {year} {2008})}\BibitemShut {NoStop}%
\bibitem [{\citenamefont {Bimonte}(2009)}]{Bimonte_plates}%
  \BibitemOpen
  \bibfield  {author} {\bibinfo {author} {\bibfnamefont {G.}~\bibnamefont
  {Bimonte}},\ }\bibfield  {title} {\bibinfo {title} {Scattering approach to
  {C}asimir forces and radiative heat transfer for nanostructured surfaces out
  of thermal equilibrium},\ }\href {https://doi.org/10.1103/PhysRevA.80.042102}
  {\bibfield  {journal} {\bibinfo  {journal} {Phys. Rev. A}\ }\textbf {\bibinfo
  {volume} {80}},\ \bibinfo {pages} {042102} (\bibinfo {year}
  {2009})}\BibitemShut {NoStop}%
\bibitem [{\citenamefont {Golyk}\ \emph {et~al.}(2012)\citenamefont {Golyk},
  \citenamefont {Kr\"uger}, \citenamefont {Reid},\ and\ \citenamefont
  {Kardar}}]{Golyk_cylinder}%
  \BibitemOpen
  \bibfield  {author} {\bibinfo {author} {\bibfnamefont {V.~A.}\ \bibnamefont
  {Golyk}}, \bibinfo {author} {\bibfnamefont {M.}~\bibnamefont {Kr\"uger}},
  \bibinfo {author} {\bibfnamefont {M.~T.~H.}\ \bibnamefont {Reid}},\ and\
  \bibinfo {author} {\bibfnamefont {M.}~\bibnamefont {Kardar}},\ }\bibfield
  {title} {\bibinfo {title} {Casimir forces between cylinders at different
  temperatures},\ }\href {https://doi.org/10.1103/PhysRevD.85.065011}
  {\bibfield  {journal} {\bibinfo  {journal} {Phys. Rev. D}\ }\textbf {\bibinfo
  {volume} {85}},\ \bibinfo {pages} {065011} (\bibinfo {year}
  {2012})}\BibitemShut {NoStop}%
\bibitem [{\citenamefont {Krüger}\ \emph {et~al.}(2011)\citenamefont
  {Krüger}, \citenamefont {Emig}, \citenamefont {Bimonte},\ and\ \citenamefont
  {Kardar}}]{Kruger_spheres}%
  \BibitemOpen
  \bibfield  {author} {\bibinfo {author} {\bibfnamefont {M.}~\bibnamefont
  {Krüger}}, \bibinfo {author} {\bibfnamefont {T.}~\bibnamefont {Emig}},
  \bibinfo {author} {\bibfnamefont {G.}~\bibnamefont {Bimonte}},\ and\ \bibinfo
  {author} {\bibfnamefont {M.}~\bibnamefont {Kardar}},\ }\bibfield  {title}
  {\bibinfo {title} {Non-equilibrium {C}asimir forces: Spheres and
  sphere-plate},\ }\href {https://doi.org/10.1209/0295-5075/95/21002}
  {\bibfield  {journal} {\bibinfo  {journal} {Europhys. Lett.}\ }\textbf
  {\bibinfo {volume} {95}},\ \bibinfo {pages} {21002} (\bibinfo {year}
  {2011})}\BibitemShut {NoStop}%
\bibitem [{\citenamefont {Polimeridis}\ \emph {et~al.}(2015)\citenamefont
  {Polimeridis}, \citenamefont {Reid}, \citenamefont {Jin}, \citenamefont
  {Johnson}, \citenamefont {White},\ and\ \citenamefont
  {Rodriguez}}]{Polimeridis_HeattransferNumerical}%
  \BibitemOpen
  \bibfield  {author} {\bibinfo {author} {\bibfnamefont {A.~G.}\ \bibnamefont
  {Polimeridis}}, \bibinfo {author} {\bibfnamefont {M.~T.~H.}\ \bibnamefont
  {Reid}}, \bibinfo {author} {\bibfnamefont {W.}~\bibnamefont {Jin}}, \bibinfo
  {author} {\bibfnamefont {S.~G.}\ \bibnamefont {Johnson}}, \bibinfo {author}
  {\bibfnamefont {J.~K.}\ \bibnamefont {White}},\ and\ \bibinfo {author}
  {\bibfnamefont {A.~W.}\ \bibnamefont {Rodriguez}},\ }\bibfield  {title}
  {\bibinfo {title} {Fluctuating volume-current formulation of electromagnetic
  fluctuations in inhomogeneous media: Incandescence and luminescence in
  arbitrary geometries},\ }\href {https://doi.org/10.1103/PhysRevB.92.134202}
  {\bibfield  {journal} {\bibinfo  {journal} {Phys. Rev. B}\ }\textbf {\bibinfo
  {volume} {92}},\ \bibinfo {pages} {134202} (\bibinfo {year}
  {2015})}\BibitemShut {NoStop}%
\bibitem [{\citenamefont {McCauley}\ \emph {et~al.}(2012)\citenamefont
  {McCauley}, \citenamefont {Reid}, \citenamefont {Kr\"uger},\ and\
  \citenamefont {Johnson}}]{McCauley_numeric}%
  \BibitemOpen
  \bibfield  {author} {\bibinfo {author} {\bibfnamefont {A.~P.}\ \bibnamefont
  {McCauley}}, \bibinfo {author} {\bibfnamefont {M.~T.~H.}\ \bibnamefont
  {Reid}}, \bibinfo {author} {\bibfnamefont {M.}~\bibnamefont {Kr\"uger}},\
  and\ \bibinfo {author} {\bibfnamefont {S.~G.}\ \bibnamefont {Johnson}},\
  }\bibfield  {title} {\bibinfo {title} {Modeling near-field radiative heat
  transfer from sharp objects using a general three-dimensional numerical
  scattering technique},\ }\href {https://doi.org/10.1103/PhysRevB.85.165104}
  {\bibfield  {journal} {\bibinfo  {journal} {Phys. Rev. B}\ }\textbf {\bibinfo
  {volume} {85}},\ \bibinfo {pages} {165104} (\bibinfo {year}
  {2012})}\BibitemShut {NoStop}%
\bibitem [{\citenamefont {Rodriguez}\ \emph
  {et~al.}(2012{\natexlab{b}})\citenamefont {Rodriguez}, \citenamefont {Reid},\
  and\ \citenamefont {Johnson}}]{Rodriguez_numericalArbitrary}%
  \BibitemOpen
  \bibfield  {author} {\bibinfo {author} {\bibfnamefont {A.~W.}\ \bibnamefont
  {Rodriguez}}, \bibinfo {author} {\bibfnamefont {M.~T.~H.}\ \bibnamefont
  {Reid}},\ and\ \bibinfo {author} {\bibfnamefont {S.~G.}\ \bibnamefont
  {Johnson}},\ }\bibfield  {title} {\bibinfo {title}
  {Fluctuating-surface-current formulation of radiative heat transfer for
  arbitrary geometries},\ }\href {https://doi.org/10.1103/PhysRevB.86.220302}
  {\bibfield  {journal} {\bibinfo  {journal} {Phys. Rev. B}\ }\textbf {\bibinfo
  {volume} {86}},\ \bibinfo {pages} {220302} (\bibinfo {year}
  {2012}{\natexlab{b}})}\BibitemShut {NoStop}%
\bibitem [{\citenamefont {Rodriguez}\ \emph {et~al.}(2013)\citenamefont
  {Rodriguez}, \citenamefont {Reid},\ and\ \citenamefont
  {Johnson}}]{Rodriguez_numerical}%
  \BibitemOpen
  \bibfield  {author} {\bibinfo {author} {\bibfnamefont {A.~W.}\ \bibnamefont
  {Rodriguez}}, \bibinfo {author} {\bibfnamefont {M.~T.~H.}\ \bibnamefont
  {Reid}},\ and\ \bibinfo {author} {\bibfnamefont {S.~G.}\ \bibnamefont
  {Johnson}},\ }\bibfield  {title} {\bibinfo {title}
  {Fluctuating-surface-current formulation of radiative heat transfer: Theory
  and applications},\ }\href {https://doi.org/10.1103/PhysRevB.88.054305}
  {\bibfield  {journal} {\bibinfo  {journal} {Phys. Rev. B}\ }\textbf {\bibinfo
  {volume} {88}},\ \bibinfo {pages} {054305} (\bibinfo {year}
  {2013})}\BibitemShut {NoStop}%
\bibitem [{\citenamefont {Narayanaswamy}\ and\ \citenamefont
  {Chen}(2008{\natexlab{b}})}]{dipole_sphere1}%
  \BibitemOpen
  \bibfield  {author} {\bibinfo {author} {\bibfnamefont {A.}~\bibnamefont
  {Narayanaswamy}}\ and\ \bibinfo {author} {\bibfnamefont {G.}~\bibnamefont
  {Chen}},\ }\bibfield  {title} {\bibinfo {title} {Thermal near-field radiative
  transfer between two spheres},\ }\href
  {https://doi.org/10.1103/PhysRevB.77.075125} {\bibfield  {journal} {\bibinfo
  {journal} {Phys. Rev. B}\ }\textbf {\bibinfo {volume} {77}},\ \bibinfo
  {pages} {075125} (\bibinfo {year} {2008}{\natexlab{b}})}\BibitemShut
  {NoStop}%
\bibitem [{\citenamefont {Shen}\ \emph {et~al.}(2009)\citenamefont {Shen},
  \citenamefont {Narayanaswamy},\ and\ \citenamefont {Chen}}]{HT_exp1}%
  \BibitemOpen
  \bibfield  {author} {\bibinfo {author} {\bibfnamefont {S.}~\bibnamefont
  {Shen}}, \bibinfo {author} {\bibfnamefont {A.}~\bibnamefont
  {Narayanaswamy}},\ and\ \bibinfo {author} {\bibfnamefont {G.}~\bibnamefont
  {Chen}},\ }\bibfield  {title} {\bibinfo {title} {Surface phonon polaritons
  mediated energy transfer between nanoscale gaps},\ }\href
  {https://doi.org/10.1021/nl901208v} {\bibfield  {journal} {\bibinfo
  {journal} {Nano Lett.}\ }\textbf {\bibinfo {volume} {9}},\ \bibinfo {pages}
  {2909} (\bibinfo {year} {2009})}\BibitemShut {NoStop}%
\bibitem [{\citenamefont {Hargreaves}(1969)}]{HT_exp2}%
  \BibitemOpen
  \bibfield  {author} {\bibinfo {author} {\bibfnamefont {C.}~\bibnamefont
  {Hargreaves}},\ }\bibfield  {title} {\bibinfo {title} {Anomalous radiative
  transfer between closely-spaced bodies},\ }\href
  {https://doi.org/10.1016/0375-9601(69)90264-3} {\bibfield  {journal}
  {\bibinfo  {journal} {Phys. Lett. A}\ }\textbf {\bibinfo {volume} {30}},\
  \bibinfo {pages} {491} (\bibinfo {year} {1969})}\BibitemShut {NoStop}%
\bibitem [{\citenamefont {Kittel}\ \emph {et~al.}(2005)\citenamefont {Kittel},
  \citenamefont {M\"uller-Hirsch}, \citenamefont {Parisi}, \citenamefont
  {Biehs}, \citenamefont {Reddig},\ and\ \citenamefont {Holthaus}}]{HT_exp3}%
  \BibitemOpen
  \bibfield  {author} {\bibinfo {author} {\bibfnamefont {A.}~\bibnamefont
  {Kittel}}, \bibinfo {author} {\bibfnamefont {W.}~\bibnamefont
  {M\"uller-Hirsch}}, \bibinfo {author} {\bibfnamefont {J.}~\bibnamefont
  {Parisi}}, \bibinfo {author} {\bibfnamefont {S.-A.}\ \bibnamefont {Biehs}},
  \bibinfo {author} {\bibfnamefont {D.}~\bibnamefont {Reddig}},\ and\ \bibinfo
  {author} {\bibfnamefont {M.}~\bibnamefont {Holthaus}},\ }\bibfield  {title}
  {\bibinfo {title} {Near-field heat transfer in a scanning thermal
  microscope},\ }\href {https://doi.org/10.1103/PhysRevLett.95.224301}
  {\bibfield  {journal} {\bibinfo  {journal} {Phys. Rev. Lett.}\ }\textbf
  {\bibinfo {volume} {95}},\ \bibinfo {pages} {224301} (\bibinfo {year}
  {2005})}\BibitemShut {NoStop}%
\bibitem [{\citenamefont {Kajihara}\ \emph {et~al.}(2011)\citenamefont
  {Kajihara}, \citenamefont {Kosaka},\ and\ \citenamefont
  {Komiyama}}]{HT_exp4}%
  \BibitemOpen
  \bibfield  {author} {\bibinfo {author} {\bibfnamefont {Y.}~\bibnamefont
  {Kajihara}}, \bibinfo {author} {\bibfnamefont {K.}~\bibnamefont {Kosaka}},\
  and\ \bibinfo {author} {\bibfnamefont {S.}~\bibnamefont {Komiyama}},\
  }\bibfield  {title} {\bibinfo {title} {Thermally excited near-field radiation
  and far-field interference},\ }\href {https://doi.org/10.1364/OE.19.007695}
  {\bibfield  {journal} {\bibinfo  {journal} {Opt. Express}\ }\textbf {\bibinfo
  {volume} {19}},\ \bibinfo {pages} {7695} (\bibinfo {year}
  {2011})}\BibitemShut {NoStop}%
\bibitem [{\citenamefont {Obrecht}\ \emph {et~al.}(2007)\citenamefont
  {Obrecht}, \citenamefont {Wild}, \citenamefont {Antezza}, \citenamefont
  {Pitaevskii}, \citenamefont {Stringari},\ and\ \citenamefont
  {Cornell}}]{exp_neq}%
  \BibitemOpen
  \bibfield  {author} {\bibinfo {author} {\bibfnamefont {J.~M.}\ \bibnamefont
  {Obrecht}}, \bibinfo {author} {\bibfnamefont {R.~J.}\ \bibnamefont {Wild}},
  \bibinfo {author} {\bibfnamefont {M.}~\bibnamefont {Antezza}}, \bibinfo
  {author} {\bibfnamefont {L.~P.}\ \bibnamefont {Pitaevskii}}, \bibinfo
  {author} {\bibfnamefont {S.}~\bibnamefont {Stringari}},\ and\ \bibinfo
  {author} {\bibfnamefont {E.~A.}\ \bibnamefont {Cornell}},\ }\bibfield
  {title} {\bibinfo {title} {Measurement of the temperature dependence of the
  {C}asimir-{P}older force},\ }\href
  {https://doi.org/10.1103/PhysRevLett.98.063201} {\bibfield  {journal}
  {\bibinfo  {journal} {Phys. Rev. Lett.}\ }\textbf {\bibinfo {volume} {98}},\
  \bibinfo {pages} {063201} (\bibinfo {year} {2007})}\BibitemShut {NoStop}%
\bibitem [{\citenamefont {Pascale}\ \emph {et~al.}(2023)\citenamefont
  {Pascale}, \citenamefont {Giteau},\ and\ \citenamefont
  {Papadakis}}]{review_HT_1}%
  \BibitemOpen
  \bibfield  {author} {\bibinfo {author} {\bibfnamefont {M.}~\bibnamefont
  {Pascale}}, \bibinfo {author} {\bibfnamefont {M.}~\bibnamefont {Giteau}},\
  and\ \bibinfo {author} {\bibfnamefont {G.~T.}\ \bibnamefont {Papadakis}},\
  }\bibfield  {title} {\bibinfo {title} {Perspective on near-field radiative
  heat transfer},\ }\href {https://doi.org/10.1063/5.0142651} {\bibfield
  {journal} {\bibinfo  {journal} {Appl. Phys. Lett.}\ }\textbf {\bibinfo
  {volume} {122}},\ \bibinfo {pages} {100501} (\bibinfo {year}
  {2023})}\BibitemShut {NoStop}%
\bibitem [{\citenamefont {V\'azquez-Lozano}\ and\ \citenamefont
  {Liberal}(2024)}]{review_HT_2}%
  \BibitemOpen
  \bibfield  {author} {\bibinfo {author} {\bibfnamefont {J.~E.}\ \bibnamefont
  {V\'azquez-Lozano}}\ and\ \bibinfo {author} {\bibfnamefont {I.}~\bibnamefont
  {Liberal}},\ }\bibfield  {title} {\bibinfo {title} {Review on the scientific
  and technological breakthroughs in thermal emission engineering},\ }\href
  {https://doi.org/10.1021/acsaom.4c00030} {\bibfield  {journal} {\bibinfo
  {journal} {ACS Appl. Opt. Mater.}\ }\textbf {\bibinfo {volume} {2}},\
  \bibinfo {pages} {898} (\bibinfo {year} {2024})}\BibitemShut {NoStop}%
\bibitem [{\citenamefont {Biehs}\ \emph {et~al.}(2021)\citenamefont {Biehs},
  \citenamefont {Messina}, \citenamefont {Venkataram}, \citenamefont
  {Rodriguez}, \citenamefont {Cuevas},\ and\ \citenamefont
  {Ben-Abdallah}}]{Biehs_review}%
  \BibitemOpen
  \bibfield  {author} {\bibinfo {author} {\bibfnamefont {S.-A.}\ \bibnamefont
  {Biehs}}, \bibinfo {author} {\bibfnamefont {R.}~\bibnamefont {Messina}},
  \bibinfo {author} {\bibfnamefont {P.~S.}\ \bibnamefont {Venkataram}},
  \bibinfo {author} {\bibfnamefont {A.~W.}\ \bibnamefont {Rodriguez}}, \bibinfo
  {author} {\bibfnamefont {J.~C.}\ \bibnamefont {Cuevas}},\ and\ \bibinfo
  {author} {\bibfnamefont {P.}~\bibnamefont {Ben-Abdallah}},\ }\bibfield
  {title} {\bibinfo {title} {Near-field radiative heat transfer in many-body
  systems},\ }\href {https://doi.org/10.1103/RevModPhys.93.025009} {\bibfield
  {journal} {\bibinfo  {journal} {Rev. Mod. Phys.}\ }\textbf {\bibinfo {volume}
  {93}},\ \bibinfo {pages} {025009} (\bibinfo {year} {2021})}\BibitemShut
  {NoStop}%
\bibitem [{\citenamefont {Ben-Abdallah}(2016)}]{Abdallah_Hall}%
  \BibitemOpen
  \bibfield  {author} {\bibinfo {author} {\bibfnamefont {P.}~\bibnamefont
  {Ben-Abdallah}},\ }\bibfield  {title} {\bibinfo {title} {Photon thermal
  {H}all effect},\ }\href {https://doi.org/10.1103/PhysRevLett.116.084301}
  {\bibfield  {journal} {\bibinfo  {journal} {Phys. Rev. Lett.}\ }\textbf
  {\bibinfo {volume} {116}},\ \bibinfo {pages} {084301} (\bibinfo {year}
  {2016})}\BibitemShut {NoStop}%
\bibitem [{\citenamefont {{Abraham Ekeroth}}\ \emph {et~al.}(2018)\citenamefont
  {{Abraham Ekeroth}}, \citenamefont {Ben-Abdallah}, \citenamefont {Cuevas},\
  and\ \citenamefont {Garc\'ia-Mart\'in}}]{Ekeroth_nanoparticles}%
  \BibitemOpen
  \bibfield  {author} {\bibinfo {author} {\bibfnamefont {R.~M.}\ \bibnamefont
  {{Abraham Ekeroth}}}, \bibinfo {author} {\bibfnamefont {P.}~\bibnamefont
  {Ben-Abdallah}}, \bibinfo {author} {\bibfnamefont {J.~C.}\ \bibnamefont
  {Cuevas}},\ and\ \bibinfo {author} {\bibfnamefont {A.}~\bibnamefont
  {Garc\'ia-Mart\'in}},\ }\bibfield  {title} {\bibinfo {title} {Anisotropic
  thermal magnetoresistance for an active control of radiative heat transfer},\
  }\href {https://doi.org/10.1021/acsphotonics.7b01223} {\bibfield  {journal}
  {\bibinfo  {journal} {ACS Photonics}\ }\textbf {\bibinfo {volume} {5}},\
  \bibinfo {pages} {705} (\bibinfo {year} {2018})}\BibitemShut {NoStop}%
\bibitem [{\citenamefont {Zhu}\ and\ \citenamefont
  {Fan}(2016)}]{Zhu_epsilon_nonreci}%
  \BibitemOpen
  \bibfield  {author} {\bibinfo {author} {\bibfnamefont {L.}~\bibnamefont
  {Zhu}}\ and\ \bibinfo {author} {\bibfnamefont {S.}~\bibnamefont {Fan}},\
  }\bibfield  {title} {\bibinfo {title} {Persistent directional current at
  equilibrium in nonreciprocal many-body near field electromagnetic heat
  transfer},\ }\href {https://doi.org/10.1103/PhysRevLett.117.134303}
  {\bibfield  {journal} {\bibinfo  {journal} {Phys. Rev. Lett.}\ }\textbf
  {\bibinfo {volume} {117}},\ \bibinfo {pages} {134303} (\bibinfo {year}
  {2016})}\BibitemShut {NoStop}%
\bibitem [{\citenamefont {Gelbwaser-Klimovsky}\ \emph
  {et~al.}(2021)\citenamefont {Gelbwaser-Klimovsky}, \citenamefont {Graham},
  \citenamefont {Kardar},\ and\ \citenamefont
  {Kr\"uger}}]{Gelbwaser_nonreciprocalPlate}%
  \BibitemOpen
  \bibfield  {author} {\bibinfo {author} {\bibfnamefont {D.}~\bibnamefont
  {Gelbwaser-Klimovsky}}, \bibinfo {author} {\bibfnamefont {N.}~\bibnamefont
  {Graham}}, \bibinfo {author} {\bibfnamefont {M.}~\bibnamefont {Kardar}},\
  and\ \bibinfo {author} {\bibfnamefont {M.}~\bibnamefont {Kr\"uger}},\
  }\bibfield  {title} {\bibinfo {title} {Near field propulsion forces from
  nonreciprocal media},\ }\href
  {https://doi.org/10.1103/PhysRevLett.126.170401} {\bibfield  {journal}
  {\bibinfo  {journal} {Phys. Rev. Lett.}\ }\textbf {\bibinfo {volume} {126}},\
  \bibinfo {pages} {170401} (\bibinfo {year} {2021})}\BibitemShut {NoStop}%
\bibitem [{\citenamefont {Khandekar}\ \emph {et~al.}(2021)\citenamefont
  {Khandekar}, \citenamefont {Buddhiraju}, \citenamefont {Wilkinson},
  \citenamefont {Gimzewski}, \citenamefont {Rodriguez}, \citenamefont {Chase},\
  and\ \citenamefont {Fan}}]{Khandekar_nonreciprocalPlate}%
  \BibitemOpen
  \bibfield  {author} {\bibinfo {author} {\bibfnamefont {C.}~\bibnamefont
  {Khandekar}}, \bibinfo {author} {\bibfnamefont {S.}~\bibnamefont
  {Buddhiraju}}, \bibinfo {author} {\bibfnamefont {P.~R.}\ \bibnamefont
  {Wilkinson}}, \bibinfo {author} {\bibfnamefont {J.~K.}\ \bibnamefont
  {Gimzewski}}, \bibinfo {author} {\bibfnamefont {A.~W.}\ \bibnamefont
  {Rodriguez}}, \bibinfo {author} {\bibfnamefont {C.}~\bibnamefont {Chase}},\
  and\ \bibinfo {author} {\bibfnamefont {S.}~\bibnamefont {Fan}},\ }\bibfield
  {title} {\bibinfo {title} {Nonequilibrium lateral force and torque by
  thermally excited nonreciprocal surface electromagnetic waves},\ }\href
  {https://doi.org/10.1103/PhysRevB.104.245433} {\bibfield  {journal} {\bibinfo
   {journal} {Phys. Rev. B}\ }\textbf {\bibinfo {volume} {104}},\ \bibinfo
  {pages} {245433} (\bibinfo {year} {2021})}\BibitemShut {NoStop}%
\bibitem [{\citenamefont {Guo}\ and\ \citenamefont
  {Fan}(2021)}]{gyrodparticle}%
  \BibitemOpen
  \bibfield  {author} {\bibinfo {author} {\bibfnamefont {Y.}~\bibnamefont
  {Guo}}\ and\ \bibinfo {author} {\bibfnamefont {S.}~\bibnamefont {Fan}},\
  }\bibfield  {title} {\bibinfo {title} {Single gyrotropic particle as a heat
  engine},\ }\href {https://doi.org/10.1021/acsphotonics.0c01920} {\bibfield
  {journal} {\bibinfo  {journal} {ACS Photonics}\ }\textbf {\bibinfo {volume}
  {8}},\ \bibinfo {pages} {1623} (\bibinfo {year} {2021})}\BibitemShut
  {NoStop}%
\bibitem [{\citenamefont {Milton}\ \emph {et~al.}(2023)\citenamefont {Milton},
  \citenamefont {Guo}, \citenamefont {Kennedy}, \citenamefont {Pourtolami},\
  and\ \citenamefont {DelCol}}]{Milton_propulsion_force}%
  \BibitemOpen
  \bibfield  {author} {\bibinfo {author} {\bibfnamefont {K.~A.}\ \bibnamefont
  {Milton}}, \bibinfo {author} {\bibfnamefont {X.}~\bibnamefont {Guo}},
  \bibinfo {author} {\bibfnamefont {G.}~\bibnamefont {Kennedy}}, \bibinfo
  {author} {\bibfnamefont {N.}~\bibnamefont {Pourtolami}},\ and\ \bibinfo
  {author} {\bibfnamefont {D.~M.}\ \bibnamefont {DelCol}},\ }\bibfield  {title}
  {\bibinfo {title} {Vacuum torque, propulsive forces, and anomalous tangential
  forces: Effects of nonreciprocal media out of thermal equilibrium},\ }\href
  {https://doi.org/10.1103/PhysRevA.108.022809} {\bibfield  {journal} {\bibinfo
   {journal} {Phys. Rev. A}\ }\textbf {\bibinfo {volume} {108}},\ \bibinfo
  {pages} {022809} (\bibinfo {year} {2023})}\BibitemShut {NoStop}%
\bibitem [{\citenamefont {M\"uller}\ and\ \citenamefont
  {Kr\"uger}(2016)}]{Muller_ellipsoid}%
  \BibitemOpen
  \bibfield  {author} {\bibinfo {author} {\bibfnamefont {B.}~\bibnamefont
  {M\"uller}}\ and\ \bibinfo {author} {\bibfnamefont {M.}~\bibnamefont
  {Kr\"uger}},\ }\bibfield  {title} {\bibinfo {title} {Anisotropic particles
  near surfaces: Propulsion force and friction},\ }\href
  {https://doi.org/10.1103/PhysRevA.93.032511} {\bibfield  {journal} {\bibinfo
  {journal} {Phys. Rev. A}\ }\textbf {\bibinfo {volume} {93}},\ \bibinfo
  {pages} {032511} (\bibinfo {year} {2016})}\BibitemShut {NoStop}%
\bibitem [{\citenamefont {Reid}\ \emph {et~al.}()\citenamefont {Reid},
  \citenamefont {Miller}, \citenamefont {Polimeridis}, \citenamefont
  {Rodriguez}, \citenamefont {Tomlinson},\ and\ \citenamefont
  {Johnson}}]{reid2017}%
  \BibitemOpen
  \bibfield  {author} {\bibinfo {author} {\bibfnamefont {M.~T.~H.}\
  \bibnamefont {Reid}}, \bibinfo {author} {\bibfnamefont {O.~D.}\ \bibnamefont
  {Miller}}, \bibinfo {author} {\bibfnamefont {A.~G.}\ \bibnamefont
  {Polimeridis}}, \bibinfo {author} {\bibfnamefont {A.~W.}\ \bibnamefont
  {Rodriguez}}, \bibinfo {author} {\bibfnamefont {E.~M.}\ \bibnamefont
  {Tomlinson}},\ and\ \bibinfo {author} {\bibfnamefont {S.~G.}\ \bibnamefont
  {Johnson}},\ }\href@noop {} {\bibinfo {title} {Photon torpedoes and {R}ytov
  pinwheels: Integral-equation modeling of non-equilibrium fluctuation-induced
  forces and torques on nanoparticles}},\ \Eprint
  {https://arxiv.org/abs/1708.01985} {arXiv:1708.01985} \BibitemShut {NoStop}%
\bibitem [{\citenamefont {Strekha}\ \emph {et~al.}(2024)\citenamefont
  {Strekha}, \citenamefont {Kr\"uger},\ and\ \citenamefont
  {Rodriguez}}]{strekha_eq}%
  \BibitemOpen
  \bibfield  {author} {\bibinfo {author} {\bibfnamefont {B.}~\bibnamefont
  {Strekha}}, \bibinfo {author} {\bibfnamefont {M.}~\bibnamefont {Kr\"uger}},\
  and\ \bibinfo {author} {\bibfnamefont {A.~W.}\ \bibnamefont {Rodriguez}},\
  }\bibfield  {title} {\bibinfo {title} {Trace expressions and associated
  limits for equilibrium {C}asimir torque},\ }\href
  {https://doi.org/10.1103/PhysRevA.109.012813} {\bibfield  {journal} {\bibinfo
   {journal} {Phys. Rev. A}\ }\textbf {\bibinfo {volume} {109}},\ \bibinfo
  {pages} {012813} (\bibinfo {year} {2024})}\BibitemShut {NoStop}%
\bibitem [{\citenamefont {Fan}\ \emph {et~al.}(2020)\citenamefont {Fan},
  \citenamefont {Guo}, \citenamefont {Papadakis}, \citenamefont {Zhao},
  \citenamefont {Zhao}, \citenamefont {Buddhiraju}, \citenamefont {Orenstein},\
  and\ \citenamefont {Fan}}]{Fan2020}%
  \BibitemOpen
  \bibfield  {author} {\bibinfo {author} {\bibfnamefont {L.}~\bibnamefont
  {Fan}}, \bibinfo {author} {\bibfnamefont {Y.}~\bibnamefont {Guo}}, \bibinfo
  {author} {\bibfnamefont {G.~T.}\ \bibnamefont {Papadakis}}, \bibinfo {author}
  {\bibfnamefont {B.}~\bibnamefont {Zhao}}, \bibinfo {author} {\bibfnamefont
  {Z.}~\bibnamefont {Zhao}}, \bibinfo {author} {\bibfnamefont {S.}~\bibnamefont
  {Buddhiraju}}, \bibinfo {author} {\bibfnamefont {M.}~\bibnamefont
  {Orenstein}},\ and\ \bibinfo {author} {\bibfnamefont {S.}~\bibnamefont
  {Fan}},\ }\bibfield  {title} {\bibinfo {title} {Nonreciprocal radiative heat
  transfer between two planar bodies},\ }\href
  {https://doi.org/10.1103/PhysRevB.101.085407} {\bibfield  {journal} {\bibinfo
   {journal} {Phys. Rev. B}\ }\textbf {\bibinfo {volume} {101}},\ \bibinfo
  {pages} {085407} (\bibinfo {year} {2020})}\BibitemShut {NoStop}%
\bibitem [{\citenamefont {Soo}\ and\ \citenamefont {Kr\"uger}(2018)}]{soo}%
  \BibitemOpen
  \bibfield  {author} {\bibinfo {author} {\bibfnamefont {H.}~\bibnamefont
  {Soo}}\ and\ \bibinfo {author} {\bibfnamefont {M.}~\bibnamefont {Kr\"uger}},\
  }\bibfield  {title} {\bibinfo {title} {Fluctuational electrodynamics for
  nonlinear materials in and out of thermal equilibrium},\ }\href
  {https://doi.org/10.1103/PhysRevB.97.045412} {\bibfield  {journal} {\bibinfo
  {journal} {Phys. Rev. B}\ }\textbf {\bibinfo {volume} {97}},\ \bibinfo
  {pages} {045412} (\bibinfo {year} {2018})}\BibitemShut {NoStop}%
\bibitem [{\citenamefont {Kr\"uger}\ \emph {et~al.}(2024)\citenamefont
  {Kr\"uger}, \citenamefont {Asheichyk}, \citenamefont {Kardar},\ and\
  \citenamefont {Golestanian}}]{Kruger_conductivity}%
  \BibitemOpen
  \bibfield  {author} {\bibinfo {author} {\bibfnamefont {M.}~\bibnamefont
  {Kr\"uger}}, \bibinfo {author} {\bibfnamefont {K.}~\bibnamefont {Asheichyk}},
  \bibinfo {author} {\bibfnamefont {M.}~\bibnamefont {Kardar}},\ and\ \bibinfo
  {author} {\bibfnamefont {R.}~\bibnamefont {Golestanian}},\ }\bibfield
  {title} {\bibinfo {title} {Scale-dependent heat transport in dissipative
  media via electromagnetic fluctuations},\ }\href
  {https://doi.org/10.1103/PhysRevLett.132.106903} {\bibfield  {journal}
  {\bibinfo  {journal} {Phys. Rev. Lett.}\ }\textbf {\bibinfo {volume} {132}},\
  \bibinfo {pages} {106903} (\bibinfo {year} {2024})}\BibitemShut {NoStop}%
\bibitem [{Note1()}]{Note1}%
  \BibitemOpen
  \bibinfo {note} {We added a minus sign to $ H_2^{(2)}$ with respect to
  Ref.~\cite {TraceFormulas} for it to be nonnegative}\BibitemShut {NoStop}%
\bibitem [{\citenamefont {Asheichyk}\ \emph {et~al.}(2017)\citenamefont
  {Asheichyk}, \citenamefont {M\"uller},\ and\ \citenamefont
  {Kr\"uger}}]{Asheichyk_heatradiation}%
  \BibitemOpen
  \bibfield  {author} {\bibinfo {author} {\bibfnamefont {K.}~\bibnamefont
  {Asheichyk}}, \bibinfo {author} {\bibfnamefont {B.}~\bibnamefont
  {M\"uller}},\ and\ \bibinfo {author} {\bibfnamefont {M.}~\bibnamefont
  {Kr\"uger}},\ }\bibfield  {title} {\bibinfo {title} {Heat radiation and
  transfer for point particles in arbitrary geometries},\ }\href
  {https://doi.org/10.1103/PhysRevB.96.155402} {\bibfield  {journal} {\bibinfo
  {journal} {Phys. Rev. B}\ }\textbf {\bibinfo {volume} {96}},\ \bibinfo
  {pages} {155402} (\bibinfo {year} {2017})}\BibitemShut {NoStop}%
\bibitem [{\citenamefont {Herz}\ and\ \citenamefont {Biehs}(2019)}]{Herz_2019}%
  \BibitemOpen
  \bibfield  {author} {\bibinfo {author} {\bibfnamefont {F.}~\bibnamefont
  {Herz}}\ and\ \bibinfo {author} {\bibfnamefont {S.-A.}\ \bibnamefont
  {Biehs}},\ }\bibfield  {title} {\bibinfo {title} {Green-{K}ubo relation for
  thermal radiation in non-reciprocal systems},\ }\href
  {https://doi.org/10.1209/0295-5075/127/44001} {\bibfield  {journal} {\bibinfo
   {journal} {Europhys. Lett.}\ }\textbf {\bibinfo {volume} {127}},\ \bibinfo
  {pages} {44001} (\bibinfo {year} {2019})}\BibitemShut {NoStop}%
\bibitem [{\citenamefont {Gelbwaser-Klimovsky}\ \emph
  {et~al.}(2022)\citenamefont {Gelbwaser-Klimovsky}, \citenamefont {Graham},
  \citenamefont {Kardar},\ and\ \citenamefont
  {Kr\"uger}}]{Gelbwaser_nonreciprocalEq}%
  \BibitemOpen
  \bibfield  {author} {\bibinfo {author} {\bibfnamefont {D.}~\bibnamefont
  {Gelbwaser-Klimovsky}}, \bibinfo {author} {\bibfnamefont {N.}~\bibnamefont
  {Graham}}, \bibinfo {author} {\bibfnamefont {M.}~\bibnamefont {Kardar}},\
  and\ \bibinfo {author} {\bibfnamefont {M.}~\bibnamefont {Kr\"uger}},\
  }\bibfield  {title} {\bibinfo {title} {Equilibrium forces on nonreciprocal
  materials},\ }\href {https://doi.org/10.1103/PhysRevB.106.115106} {\bibfield
  {journal} {\bibinfo  {journal} {Phys. Rev. B}\ }\textbf {\bibinfo {volume}
  {106}},\ \bibinfo {pages} {115106} (\bibinfo {year} {2022})}\BibitemShut
  {NoStop}%
\bibitem [{\citenamefont {Ishimaru}(2017)}]{Ishimaru}%
  \BibitemOpen
  \bibfield  {author} {\bibinfo {author} {\bibfnamefont {A.}~\bibnamefont
  {Ishimaru}},\ }\href@noop {} {\emph {\bibinfo {title} {Electromagnetic Wave
  Propagation, Radiation, and Scattering: From Fundamentals to Applications}}}\
  (\bibinfo  {publisher} {John Wiley \& Sons},\ \bibinfo {year}
  {2017})\BibitemShut {NoStop}%
\bibitem [{\citenamefont {Bohren}\ and\ \citenamefont
  {Huffman}(1983)}]{Absorption_and_scatteering}%
  \BibitemOpen
  \bibfield  {author} {\bibinfo {author} {\bibfnamefont {C.~F.}\ \bibnamefont
  {Bohren}}\ and\ \bibinfo {author} {\bibfnamefont {D.~R.}\ \bibnamefont
  {Huffman}},\ }\href@noop {} {\emph {\bibinfo {title} {Absorption and
  Scattering of Light by Small Particles}}}\ (\bibinfo  {publisher} {John Wiley
  \& Sons},\ \bibinfo {year} {1983})\BibitemShut {NoStop}%
\bibitem [{\citenamefont {Milton}\ \emph {et~al.}(2010)\citenamefont {Milton},
  \citenamefont {Parashar}, \citenamefont {Wagner},\ and\ \citenamefont
  {Cavero-Peláez}}]{Milton_lateralforce_eq}%
  \BibitemOpen
  \bibfield  {author} {\bibinfo {author} {\bibfnamefont {K.~A.}\ \bibnamefont
  {Milton}}, \bibinfo {author} {\bibfnamefont {P.}~\bibnamefont {Parashar}},
  \bibinfo {author} {\bibfnamefont {J.}~\bibnamefont {Wagner}},\ and\ \bibinfo
  {author} {\bibfnamefont {I.}~\bibnamefont {Cavero-Peláez}},\ }\bibfield
  {title} {\bibinfo {title} {Multiple scattering {C}asimir force calculations:
  Layered and corrugated materials, wedges, and {C}asimir-{P}older forces},\
  }\href {https://doi.org/10.1116/1.3292607} {\bibfield  {journal} {\bibinfo
  {journal} {J. Vac. Sci. Technol. B}\ }\textbf {\bibinfo {volume} {28}},\
  \bibinfo {pages} {C4A8} (\bibinfo {year} {2010})}\BibitemShut {NoStop}%
\bibitem [{Note2()}]{Note2}%
  \BibitemOpen
  \bibinfo {note} {In Ref.~\cite {Milton_propulsion_force}, $T_1=T_{\protect
  \mathrm {env}}$. In this case, the interaction force is canceled out by the
  environment, and it is sufficient to calculate the self force, see
  Eq.~\protect \eqref {eq:total_force}.}\BibitemShut {Stop}%
\bibitem [{\citenamefont {Incardone}\ \emph {et~al.}(2014)\citenamefont
  {Incardone}, \citenamefont {Emig},\ and\ \citenamefont
  {Krüger}}]{Incardone_2014}%
  \BibitemOpen
  \bibfield  {author} {\bibinfo {author} {\bibfnamefont {R.}~\bibnamefont
  {Incardone}}, \bibinfo {author} {\bibfnamefont {T.}~\bibnamefont {Emig}},\
  and\ \bibinfo {author} {\bibfnamefont {M.}~\bibnamefont {Krüger}},\
  }\bibfield  {title} {\bibinfo {title} {Heat transfer between anisotropic
  nanoparticles: Enhancement and switching},\ }\href
  {https://doi.org/10.1209/0295-5075/106/41001} {\bibfield  {journal} {\bibinfo
   {journal} {Europhys. Lett.}\ }\textbf {\bibinfo {volume} {106}},\ \bibinfo
  {pages} {41001} (\bibinfo {year} {2014})}\BibitemShut {NoStop}%
\bibitem [{\citenamefont {Nikbakht}(2014)}]{Nikbakht_2014}%
  \BibitemOpen
  \bibfield  {author} {\bibinfo {author} {\bibfnamefont {M.}~\bibnamefont
  {Nikbakht}},\ }\bibfield  {title} {\bibinfo {title} {Radiative heat transfer
  in anisotropic many-body systems: Tuning and enhancement},\ }\href
  {https://doi.org/10.1063/1.4894622} {\bibfield  {journal} {\bibinfo
  {journal} {J. Appl. Phys.}\ }\textbf {\bibinfo {volume} {116}},\ \bibinfo
  {pages} {094307} (\bibinfo {year} {2014})}\BibitemShut {NoStop}%
\bibitem [{\citenamefont {Tsang}\ \emph {et~al.}(2000)\citenamefont {Tsang},
  \citenamefont {Kong},\ and\ \citenamefont {Ding}}]{Tsang_Mie}%
  \BibitemOpen
  \bibfield  {author} {\bibinfo {author} {\bibfnamefont {L.}~\bibnamefont
  {Tsang}}, \bibinfo {author} {\bibfnamefont {J.~A.}\ \bibnamefont {Kong}},\
  and\ \bibinfo {author} {\bibfnamefont {K.~H.}\ \bibnamefont {Ding}},\
  }\href@noop {} {\emph {\bibinfo {title} {Scattering of Electromagnetic Waves,
  Theories and Applications}}}\ (\bibinfo  {publisher} {Wiley},\ \bibinfo
  {year} {2000})\BibitemShut {NoStop}%
\bibitem [{\citenamefont {Lakhtakia}\ \emph {et~al.}(1991)\citenamefont
  {Lakhtakia}, \citenamefont {Varadan},\ and\ \citenamefont
  {Varadan}}]{Lakhtakia_alpha_magneto_optical}%
  \BibitemOpen
  \bibfield  {author} {\bibinfo {author} {\bibfnamefont {A.}~\bibnamefont
  {Lakhtakia}}, \bibinfo {author} {\bibfnamefont {V.}~\bibnamefont {Varadan}},\
  and\ \bibinfo {author} {\bibfnamefont {V.}~\bibnamefont {Varadan}},\
  }\bibfield  {title} {\bibinfo {title} {Low-frequency scattering by an
  imperfectly conducting sphere immersed in a dc magnetic field},\ }\href
  {https://doi.org/10.1007/BF01014683} {\bibfield  {journal} {\bibinfo
  {journal} {Int. J. Infrared Millim. Waves}\ }\textbf {\bibinfo {volume}
  {12}},\ \bibinfo {pages} {1253} (\bibinfo {year} {1991})}\BibitemShut
  {NoStop}%
\end{thebibliography}%
%

\end{document}